\documentclass[12pt,a4paper,twoside]{article}
\usepackage[english]{babel}
\usepackage[T1]{fontenc}
\usepackage[latin9]{inputenc}
\usepackage{xcolor}
\usepackage{pdfcolmk}
\usepackage{refstyle}
\PassOptionsToPackage{normalem}{ulem}
\usepackage{ulem}
\makeatletter

\AtBeginDocument{\providecommand\eqref[1]{\ref{eq:#1}}}
\RS@ifundefined{subsecref}
{\newref{subsec}{name = \RSsectxt}}
{}
\RS@ifundefined{thmref}
{\def\RSthmtxt{theorem~}\newref{thm}{name = \RSthmtxt}}
{}
\RS@ifundefined{lemref}
{\def\RSlemtxt{lemma~}\newref{lem}{name = \RSlemtxt}}
{}

\makeatother
\def\TheAuthor{Frank K. Wilhelm$^1$, Susanna Kirchhoff$^1$, Shai Machnes$^1$,
	Nicolas Wittler$^1$, and Dominique Sugny$^2$}                                   % Author name
\def\TheTitle{An introduction into optimal control for quantum technologies}                        % Title of contribution
\def\TheHeading{An introduction into optimal control for quantum technologies}                                % Short running title
\def\TheInstitute{$^1$Theoretical Physics, Saarland University, 66123 Saarbr\"ucken }%$^{2}$Laboratoire Interdisciplinaire Carnot de Bourgogne (ICB) }                               % Group, Institute
\def\TheAddress{$^2$Laboratoire Interdisciplinaire Carnot de Bourgogne (ICB)\newline UMR 6303 CNRS-Université Bourgogne-Franche Comté, 21078 Dijon Cedex, France}            % Place e.g. University ...

\def\ThePart{B}                                             % Part of book
\def\TheChapter{4}

\usepackage{ferienschule_20}                                  % Style file for layout

\makeindex
\usepackage{babel}

\newcommand{\myiffindex}[1]{\iffindex{#1}#1}

\begin{document}

	\MakeTitel           % Displays title, author name, table of contents, footnote etc.
	%\MakeTitel[1pt]     % use this to pack contents into one page

	\newpage

\global\long\def\ket#1{\mbox{\ensuremath{\left|#1\right\rangle }}}%

\global\long\def\bra#1{\left\langle #1\right|}%

\global\long\def\braket#1#2{\left\langle #1\vphantom{#2}\right|\left.#2\vphantom{#1}\right\rangle }%

\global\long\def\ketbra#1#2{\left|#2\vphantom{#1}\right\rangle \left\langle #1\vphantom{#2}\right|}%

\global\long\def\Braket#1#2#3{\left\langle #1\vphantom{\{#2#3\}}\right|#2\left|\vphantom{\{#1#2\}}#3\right\rangle }%

\section{Introduction}

Control is a key component in turning science into technology\cite{Glaser2015},
\cite{Acin2018}. Broadly and colloquially speaking, control looks
at providing the user / experimenter with external parameters to steer
a given dynamical system to her liking \footnote{where female attributions are the default, male is considered included}
rather than simply observing its internal dynamics. Control is in
this sense ubiquitous to modern technology. In this colloquial sense,
quantum control is transferring that idea to quantum systems and thus
contains both hard- and software of many kinds.

The ubiquity of control has given rise to the field of control theory.
This is a field of applied mathematics that looks at how to choose
said external parameters in order to drive the dynamical system to
one\textquoteright s liking. It has spawned ideas of open-loop control,
i.e., the pre-determination of controls given the laws of nature (that
were a key ingredient to, e.g., the Apollo program) as well as closed-loop-control,
interleaving of observation and adjustment as we know it in our daily
lives from thermostats. This type of optimal control theory takes
the hardware setup as a given, however ideally, these are developed
in tandem. The mathematical procedures of open-loop-control typically
involve optimizing a cost function, hence the name optimal control.

The application of optimal control is not an entirely new idea. Pioneering
applications were primarily chemistry, such as the laser control of
chemical reactions and magnetic resonance. By now, quantum optimal
control is also applied to a large spectrum of modern quantum technologies
(Quantum 2.0) \cite{Acin2018}. This implies a certain tradition of fragmentation
- quantum optimal control researchers tend to be in departments of
mathematics, chemistry, computer science, and physics and follow their
specific idiosyncrasies \cite{Glaser2015}. Modern efforts have gone very far in overcoming
this fragmentation which is fruitful in learning from each other
and respecting the different goals -- quantum control of complex reactions
does for example deal with large Hilbert spaces whereas control
in quantum computing aims at sufficiently low errors in order to meet
error correction thresholds.

In this series of lectures, we would like to introduce the audience
to quantum \myiffindex{optimal control}. The first lecture will cover basic ideas
and principles of optimal control with the goal of demystifying its
jargon. The second lecture will describe computational tools (for
computations both on paper and in a computer) for its implementation
as well as their conceptual background. The third chapter will go
through a series of popular examples from different applications of
quantum technology.

These are lectures notes. Other than a textbook, it makes a significant
difference to attend the lectures it goes with rather than use it
to self-study. Other than a review, it is not complete but rather
serves to introduce clarify the concepts of the field.
This also means that the choice of references is certainly not complete,
rather, it is the subjective choice of what the authors find most
suitable and got inspired by.

\pagebreak

\section{Elementary optimal control}

We start with classical examples of control, which lay a lot of foundations for
quantum systems.

\subsection{Optimal control of a classical harmonic oscillator \label{CH:SHO}}

In order to understand the basic concept and structure of quantum
optimal control, let us start with a simple classical example: control
of the harmonic oscillator.

The equation of motion of a harmonic oscillator driven by force $F(t)=mf(t)$
where $m$ is the mass and eigenfrequency $\Omega$ is given by
\[
\ddot{x}+\Omega^{2}x(t)=f(t).
\]

Its general solution is parameterized through the Green's function
\[
G(\tau)=\frac{\theta(\tau)}{\Omega}\sin\left(\Omega\tau\right)
\]

(where $\theta$ is the Heaviside function) as
\[
x(t)=x(0)\cos\Omega t+\frac{\dot{x}(0)}{\Omega}\sin\Omega t+\int_{0}^{t}dt^{\prime}\ \frac{\sin\Omega\left(t-t^{\prime}\right)}{\Omega}f\left(t^{\prime}\right)
\]
 Readers not familiar with Green's functions can easily verify that
this expression does indeed solve the equation of motion of the driven
oscillator.

From this we get the velocity
\[
\dot{x}(t)=\dot{x}(0)\cos\Omega t-\Omega x(0)\sin\Omega t+\int_{0}^{t}dt^{\prime}\cos\Omega(t-t^{\prime})f\left(t^{\prime}\right)
\]

Thus, imposing target values $x(T)$ and $\dot{x}(T)$ we find the
conditions
\begin{align*}
x(T)-x(0)\cos\Omega T - \frac{\dot{x}(0)}{\Omega}\sin\Omega t& =\int_{0}^{T}dt^{\prime}\ \frac{\sin\Omega\left(T-t^{\prime}\right)}{\Omega}f\left(t^{\prime}\right)\\
\dot{x}(T)-\dot{x}\left(0\right)\cos\Omega T +\Omega x(0)\sin\Omega t& =\int_{0}^{T}dt^{\prime}\,\cos\Omega(T-t^{\prime})f(t^{\prime}).
\end{align*}

These equations allow a few observations that have analogies all over
quantum control: Firstly, the control $f(t)$ is needed to push the
system away from its natural dynamics (the terms on the left)
-- it is redirecting the natural drift of the system. Secondly, there
are two constraints for a function given through an integral -- so
we can expect many solutions. As an example, let's look at the case
that we move a particle by a fixed distance $x(0)=0$ and $x(T)=X$
from rest to rest $\dot{x}(T)=\dot{x}(0)=0$. We thus need to satisfy
\[
\int_{0}^{T}dt^{~\prime}\sin\left[\Omega\left(t-t^{\prime}\right)\right]f\left(t^{\prime}\right)=\Omega X\quad\int_{0}^{T}dt^{\prime}\cos\left[\Omega(t-t^{\prime})\right]f\left(t^{\prime}\right)=0
\]
 and we can easily show that this fixes low Fourier components of
$f\left(t^{\prime}\right)$ but leaves higher ones open.

The situation changes, if we impose, e.g., an energetic constraint
to the control. This typically leads to constraints of the form
$$
\int_0^T dt\ f^2(t)=\int_{-\infty}^\infty d\omega f(\omega)f^\ast(\omega)\le A
$$
where A is the imposed maximum and we have used symmetry properties of the Fourier transform of real-valued functions. Thus,  the sum of Fourier components needs to be bounded and if the
constraint is too close, there may not even be any solution. This
is an example showing that constraints clearly influence the number
of accessible solutions and their potential performance, which is commonly seen in practice.

\subsection{Optimal control for a classical system}

The previous section hinged on having a closed-form Green's function
solution of the equation of motion, which is not always available.
This follows chapter 2.3 of Bryson and Ho \cite{bryson1975applied}.

Suppose we have a dynamical system that can be controlled by a
control parameter $u$ that enters a dynamic equation for the state
variable $x$ in the form
\begin{equation}
\dot{x}=f\left[x(t),u(t),t\right]\quad0\le t\le T\label{eq:Classic_EOM}
\end{equation}
 with a given $x(0)$. Both $x$ and $u$ can be single variables or vectors of variables.  We wish to optimize a cost function at the
end of the process $J\left[x(T),T\right].$ We recall classical Lagrangian
mechanics and introduce a \myiffindex{Lagrange multiplier} function $\lambda$ and
can thus state based on the constrained calculus of variations that
we need to find a stationary point of
\[
\bar{J}=J\left[x(T),T\right]+\int_{0}^{T}dt\ \lambda^{T}(t)\left(f\left[x(t),u(t),t\right]-\dot{x}\right)
\]
 where we have allowed for the complex of coupled equations and thus
vector-valued Lagrange multipliers.

%Nico
The introduction of the Lagrange multiplier allows for the optimization of $J$, while satisfying the equation of motion (\ref{eq:Classic_EOM}) at specified times. As such, this means $\lambda$ has to be time-dependent as well.

We introduce the associated Hamilton's
function (which has a similar mathematical origin in the calculus
of variations as the Hamiltonians of mechanics yet a very different
physical motivation)
\begin{equation}
H\left[x(t),u(t),\lambda(t),t\right]=\lambda^{T}(t)f\left[x(t),u(t),t\right]
\label{eq:hamiltonian}
\end{equation}
 and rewrite our constrained cost function by integrating the last
term by parts
\[
\bar{J}=J\left[x(T),T\right]+\lambda^{T}(T)x(T)-\lambda^{T}(0)x(0)+\int_{0}^{T}dt\ \left\{ H\left[x(t),u(t),\lambda(t),t\right]+\dot{\lambda}^{T}x(t)\right\} .
\]
%DS: example for J?
Now let's consider the variation in $\bar{J}$ based on variations
in $u(t)$ recalling that the times as well as the initial state variable
are given. We find
\[
\delta\bar{J}=\left.\left(\frac{\partial J}{\partial x}-\lambda^{T}\right)\delta x\right|_{t=T}+\left.\lambda^{T}\delta x\right|_{t=0}+\int_{0}^{T}dt\ \left[\left(\frac{\partial H}{\partial x}+\dot{\lambda}^{T}\right)\delta x+\frac{\partial H}{\partial u}\delta u\right].
\]
%Nico
Note that in general we choose the variation at the beginning to be $\delta x(0)=0$, since we know the exact initial state of the dynamics.

Now the variations of $x$ and $u$ are not independent, they are
linked by the equation of motion. Were we not to work with the Lagrange
multiplier, we would need to tediously solve the equation of motion
for different control functions and then work out how these variations
are related. Fortunately, the Lagrange multiplier method allows us
to circumvent that problem.
%Nico
Our goal is for $\delta\bar{J}$ to vanish to first order. Choosing a specific Lagrange multiplier to realize this, we finally arrive at
\begin{equation}
\dot{\lambda}^{T}=-\frac{\partial H}{\partial x}=-\lambda^{T}\frac{\partial f}{\partial x}\quad\lambda^{T}(t_{f})=\frac{\partial J}{\partial x(t_f)}.\label{eq:classical_back-propagation}
\end{equation}
These are the Euler-Lagrange equations pertaining to the system.
 That being satisfied, we are left with the total variation
\[
\delta\bar{J}=\underset{=0}{\underbrace{\lambda^{T}(0)\delta x(0)}}+\int_{0}^{T}dt\ \frac{\partial H}{\partial u}\delta u
\]

For an extremum to be reached under any variation of the control,
we need
\begin{equation}
\frac{\partial H}{\partial u}=\lambda^{T}\frac{\partial f}{\partial u}=0\quad0\le t\le T.\label{eq:classical_gradient}
\end{equation}

We have shown the ingredients to what can be formalized as the 
\myiffindex{Pontryiagin Maximum Principle} (PMP). More pragmatically, these equations give
us a recipe on how to solve the thus formulated optimal control problem
by a coupled gradient search: From a suitable initial guess for $u(t)$
\begin{enumerate}
\item Solve the equation of motion eq. (\ref{eq:Classic_EOM}) to find $x(t)$
using the initial value $x(0)$ that is part of the control problem
\item Find the Lagrange multiplier by solving eq. (\ref{eq:classical_back-propagation}).
Note that there is a definite value given at the \emph{end }time $T$,
i.e., we have a final value problem -- that is solved like an initial
value problem but propagating backwards in time. This back-propagation
is typical when we consider this cost functional.
\item With these, compute the effective gradient in eq. (\ref{eq:classical_gradient})
and update the values of $u$ following the direction of the gradient.
Adjust the step size as needed.
%DS: Here we could add that a choice at first order is:
%\delta u=\varepsilon \partial H/\partial in order to ensure that \bar{J} increases or decreases at each step of the algorithm.
\end{enumerate}
Iterating these three steps will get us to a local solution, depending on the initial conditions, if the control
landscape admits one.

\subsubsection{Example: Driven harmonic oscillator}

Let us get back to formulating these steps for the optimal control
problem of the driven harmonic oscillator described above in section \ref{CH:SHO}. We identify
the control as the dimensionless force $u\equiv f$ and write the
equation of motion as a coupled system
\begin{align*}
\frac{dx}{dt} & =\dot{x}\\
\frac{d\dot{x}}{dt} & =-\Omega^{2}x+u\\
x(0) & =0\quad\dot{x}(0)=0
\end{align*}

In order to have a differentiable performance index that forces the
particle to end at $a$ at time $T$ and in rest we can write
\begin{equation}
J=\Omega^{2}\left(x-a\right)^{2}+\dot{x}^{2}.
\label{eq:sho_index}
\end{equation}
This leads us to Hamilton's function following the prescription of \ref{eq:hamiltonian}
\[
H=\lambda_{1}\dot{x}+\lambda_{2}\left(u-\Omega^{2}x\right).
\]

So the Euler-Lagrange equations \ref{eq:classical_back-propagation} describing the Lagrange Multiplier
\begin{equation}
\dot{\lambda}_{1}=\lambda_{2}\Omega^{2}\quad\dot{\lambda}_{2}=-\lambda_{1}
\label{eq:sho_adjoint}
\end{equation}
which remarkably describes a free harmonic oscillator. It is such
interpretations that lead to the Lagrange multiplier to be called
the \myiffindex{adjoint system. }The final conditions from eq. (\ref{eq:classical_back-propagation}) are
\begin{equation}
\lambda_{1}(T)=2\Omega^{2}(x(T)-a)\quad\lambda_{2}(T)=2\dot{x}(T)
\label{eq:sho_final}
\end{equation}
which are of course both zero if the final conditions are met (thus,
for the optimal solution, the adjoint system vanishes at $T$). The gradient flow
for the control is given by eq. (\ref{eq:classical_gradient})
\[
\frac{\partial H}{\partial u}=\lambda_{2}.
\]

Again, iterating these equations will give us a suitable control.

We could guess as a first control that $u_{0}(t)=\Omega^{2}a$ (which
is the force needed to keep the particle at rest at the final position,
so at least a motivated guess) thus leading to the equation of motion
\[
\ddot{x}_{0}+\Omega^{2}(x_{0}-a)=0
\]
with the solution$x_{0}(t)=a\left(1-\cos\Omega t\right)$ and thus
$\dot{x}_{0}=a\Omega\sin\Omega T.$ This clearly does not solve the
control problem, we have from eq. (\ref{eq:sho_index})  $J=\Omega^{2}a^{2}$. In fact, the final
conditions eq. \ref{eq:sho_final} for the adjoint system are $\lambda_{1}(T)=-2\Omega^{2}a\cos\Omega T$
and $\lambda_{2}(T)=2a\Omega\sin\Omega T$ leading us, by solving eq. (\ref{eq:sho_adjoint})
\begin{align*}
\lambda_{1} & =-2\Omega^{2}a\cos\Omega T\cos\left[\Omega\left(t-T\right)\right]+2a\Omega^{2}\sin\Omega T\sin\left[\Omega\left(t-T\right)\right]\\
 & =-2a\Omega^{2}\cos\Omega t
\end{align*}

and $\lambda_{2}=-2\sin\Omega t$ . This means that the gradient suggests
introducing a resonant drive -- as we have seen from the
exact solution above.

For further treatment of the classical Harmonic oscillator, see \cite{andresen2011optimal}.

\subsection{Gradient-based optimal quantum control with the GRAPE algorithm}

These principles can be transferred to the control of quantum systems
in a straightforward way. This is easily illustrated with the 
\myiffindex{GRadient Ascent Pulse Engineering (GRAPE)} algorithm \cite{GRAPE}.

\subsubsection{State-to-state control}

We start with a simple state preparation problem. Suppose WLOG that
our system is described by a Hamiltonian
\[
\hat{H}(t)=\hat{H}_{0}+\sum_{i=1}^{n}u_{i}(t)\hat{H}_{i}.
\]
We call the time-independent part of the Hamiltonian $\hat{H}_{0}$
the \myiffindex{drift}, the fields $u_{i}$ are the controls and $\hat{H}_{i}$
are the control Hamiltonians. In atomic physics, say, $\hat{H}_{0}$
describes the energy level structure of the atom, $u_{i}$ are laser
or microwave fields and $\hat{H}_{i}$ are dipole operators describing
the different field modes including polarization. Our task is now
to start at an initial state $|\psi_{0}\rangle$ at time $t=0$ and
find controls such that we reach state $|\psi_{1}\rangle$ at time
$t=T.$ As in quantum physics the global phase is meaningless, this
corresponds to maximizing the overlap $J=\left|\left\langle \psi_{1}|\psi(T)\right\rangle \right|^{2}$.
The dynamics of our system is, of course, subject to the Schrödinger
equation
\[
i\hbar\partial_{t}|\psi(t)\rangle=\hat{H}(t)|\psi(t)\rangle.
\]

Mathematically we got ourselves a system of the exact same structure
as the previous one. We give its derivation in the form of Ref. \cite{GRAPE}.

Many practical generators for $u_{i}$ such as standard arbitrary
wave form generators (AWGs or Arbs) used in superconducting qubits
represent\footnote{but not necessarily output, as the output is typically smoothed and
filtered} the pulse in a piecewise constant fashion , so it is natural \footnote{although not always optimal, see below}
to represent the $u_{i}(t)$ in that same way: We chop the total time
into $N$ intervals of length $\delta t=T/N$ and write
\[
u_{i}(t)=u_{i}(j)\quad{\rm for}\quad(j-1)\delta t\le t<j\delta t.
\]
 This allows us to write down the formal solution of the Schrödinger
equation as
\[
\hat{U}(T)=\hat{U}_{N}\hat{U}_{N-1}\cdots\hat{U}_{2}\hat{U}_{1}
\]
with
\begin{equation}
\hat{U}_{k}=\exp\left(-\frac{i}{\hbar}\delta_{t}\left(\hat{H}_{o}+\sum_{i}u_{i}(j)\hat{H}_{i}\right)\right)\label{eq:PWC-Uk}
\end{equation}
 which we can introduce into the performance index as
\[
J=\left|\left\langle \psi_{1}|\hat{U}(T)\psi_{0}\right\rangle \right|^{2}=\left|\left\langle \psi_{1}|\hat{U}_{N}\cdots\hat{U}_{1}\psi_{0}\right\rangle \right|^{2}.
\]
 We are at liberty to move some of the factors into the adjoint state, giving
us
\[
J=\left|\left\langle U_{m+1}^{\dagger}\cdots U_{N}^{\dagger}\psi_{1}|\hat{U}_{m}\cdots\hat{U}_{1}\psi_{0}\right\rangle \right|^{2}
\]
or $J=\left|\left\langle \lambda_{m}|\rho_{m}\right\rangle \right|^{2}$
with $|\rho_{m}\rangle=\hat{U}_{m}\cdots\hat{U}_{1}|\psi_{0}\rangle$
$|\lambda_{m}\rangle=\hat{U}_{N}\cdots\hat{U}_{m+1}|\psi_{1}\rangle$. Here, the partially propagated state $|\rho_m(t)\rangle$ is overlapped with the partially back-propagated \myiffindex{adjoint state} $|\lambda_m(t)\rangle$ -- both states are overlapped at time $t_m$.  We thus
sweep the time at which we calculate the overlap based on the actual
pulse that we apply. Now the final ingredient we need is the derivative
of an exponential proven in Theorem 4.5 of \cite{Hall00} (see also \cite{Hall03})
\begin{equation}
\left.\frac{d}{dt}\right|_{t=0}e^{X+tY}=e^{X}\left\{ Y-\frac{\left[X,Y\right]}{2!}+\frac{\left[X,\left[X,Y\right]\right]}{3!}-\dots\right\} \label{eq:Deriv-of-time-slice}
\end{equation}
Both of these together allow us to determine all the gradients needed to compute an update at any time step as shown in the left column of figure \ref{fig:update}.

We can rewrite this as
\[
\left.\frac{d}{dt}\right|_{t=0}e^{X+tY}=e^{X}\int_{0}^{1}d\tau\ e^{\tau X}Ye^{-\tau X}
\]
by simple power counting. This allows us to analytically compute the
derivative of the propagator across one time step by identifying $\hat{X}=\hat{H}(t)$
(the Hamiltonian including the current values of the control) and
$\hat{Y}=\hat{H}_{i}$ , one of the control Hamiltonians. In order
to simplify the right-hand side, we define $\hat{U}_{k}(j)=\hat{U}_{k}^{j}$
(taking the exponential here simply means to stretch time and study
the integral on the right
\begin{align*}
\int_{0}^{1}d\tau\ &\hat{U}_{k}(j)\hat{H}_{i}\hat{U}_{k}^{\dagger}(j)\\
& =\int d\tau\ \left(1-i\tau\delta_{t}\hat{H}-\tau^{2}\delta_{t}^{2}\hat{H}^{2}+\dots\right)\hat{H}_{i}\left(1+i\tau\delta_{t}\hat{H}-\tau^{2}\delta_{t}^{2}\hat{H}^{2}+\dots\right)\\
 & =\int d\tau\ \left(\hat{H}_{i}-i\tau\delta_{t}\left[\hat{H},\hat{H}_{i}\right]+\dots\right)\\
 &\simeq\hat{H}_{i}
\end{align*}
where we assume that the time steps chosen are so small that the integral
over the commutator can be neglected\footnote{we will later, under the Magnus expansion, study related steps more
carefully}. A self-contained derivation is presented later in  \ref{subsec:The-gradient-of-a-matrix-exponential}. Restoring
all the units leads us to the closed gradient formula
\begin{equation}
\frac{\partial J}{\partial u_{i}(j)}=-i\delta_{t}\left\langle \lambda_{j}\left|\hat{H}_{i}\right|\rho_{j}\right\rangle \label{eq:grape_gradient_state}
\end{equation}
 meaning that we can expect, with an appropriate value of $\epsilon$
compute a gradient-based update
\begin{equation}
u_{i}(j)\mapsto u_{i}(j)+\epsilon\frac{\partial J}{\partial u_{i}(j)}\label{eq:grape_update_state}
\end{equation}

This allows us to extremalize $J$ hence to find controls that best
approximate the final state with the following algorithm. Starting
from an initial guess for the controls:
\begin{enumerate}
\item Compute the propagated initial state $|\rho_{m}\rangle=\hat{U}_{m}\cdots\hat{U}_{1}|\psi_{0}\rangle$
for all $m\le N$ by iterative matrix multiplication.
\item Compute the back-propagated final state $|\lambda_{m}\rangle=\hat{U}_{N}\cdots\hat{U}_{m+1}|\psi_{1}\rangle$
by iterative matrix multiplication
\item Compute the gradient of the performance index and update the controls
following eqs. (\ref{eq:grape_gradient_state}), (\ref{eq:grape_update_state})
\item Iterate until the value of $J$ is satisfactory or the updates are
below a threshold
\end{enumerate}
There are a lot of practical improvements that were found beyond this
which we will describe below.

One must not underestimate the importance of this analytical derivation of a gradient. Whenever a gradient is available, it greatly improves the convergence of a search specifically when going from a rough initial guess that can often be obtained by solving an approximate version of the problem at hand to a solution that has the very high precision generally demanded by quantum technologies. If a gradient is available, its analytical and exact derivation is also paramount -- numerical gradients are very hard to control numerically as they involve a small difference between two potentially large numbers. In pioneering, pre-GRAPE work \cite{Niskanen03} this was rather obvious -- even with large computational effort, only few parameters could be optimized.

\subsubsection{An alternative, direct derivation}

An alternative derivation of the variational approach to quantum optimal
control is as follows:

Let us again look at the state transfer task. We shall construct a
functional, $J$, to be maximized, and utilize Lagrange multipliers
to enforce both the intial condition and the equation of motion. We
shall parameterize our control fields, $u\left(t\right)$ using a vector
of scalar real parameters $\vec{\alpha}$.

Our aim is to maximize the overlap of the goal state $\ket{\psi_{\textrm{goal}}}$
and the state at final time $T$, $\ket{\psi\left(T\right)}$,

\begin{equation}
J_{\textrm{goal}}=\left|\braket{\psi\left(T\right)}{\psi_{\textrm{goal}}}\right|^{2}.
\end{equation}

We need to impose an initial condition, utilizing a Lagrange multiplier
$\lambda_{\textrm{init}}$

\begin{equation}
J_{\textrm{init}}=\lambda_{\textrm{init}}\left(\left|\braket{\psi\left(0\right)}{\psi_{\textrm{init}}}\right|^{2}-1\right).
\end{equation}

Next, we must guarantee the Schrödinger equation, $\left(i\hbar\partial_{t}-H\left(\bar{\alpha},t\right)\right)\ket{\psi\left(t\right)}=0$
is upheld at all times. To do that, at each point in time, $t$, we
must multiply the equation of motion by the Lagrange multiplier $\bra{\chi\left(t\right)}$,
and we must add the contributions for all points in time:
\begin{equation}
J_{\textrm{e.o.m}}=\int_{o}^{T}\Braket{\chi\left(t\right)}{i\hbar\partial_{t}-H\left(\bar{\alpha},t\right)}{\psi\left(t\right)}
\end{equation}

Note that in $J_{\textrm{e.o.m}}$, $\bra{\chi\left(t\right)}$ can
be interpreted as a conjugate state, propagating backwards in time,
as the term can be rewritten as $\braket{\left(-i\hbar\partial_{t}-H\left(\bar{\alpha},t\right)\right)\chi\left(t\right)}{\psi\left(t\right)}$.

The functional to be minimized is then
\begin{equation}
J=J_{\textrm{init}}+J_{\textrm{e.o.m}}+J_{\textrm{goal}}
\end{equation}

We then proceed in the standard variational approach, taking the gradient
of this functional with respect to $\bar{\alpha}$ and requiring

\begin{equation}
\partial_{\bar{\alpha}}J=0.
\end{equation}

\subsubsection{Synthesis of unitary gates}

We will now go to the topic of finding controls that best approximate
a quantum gate. This can be viewed as a generalization of the state
preparation problem to rotating a full basis of the Hilbert space
into a desired new basis. This first begs the question of how to find
an appropriate performance index. It can be accomplished by starting
with a distance measure between the desired and the actual final unitary
$\left\Vert \hat{U}_{{\rm target}}-\hat{U}(T)\right\Vert $. The most
common choice is based on the 2-norm
\begin{align*}
\left\Vert \hat{U}_{{\rm target}}-\hat{U}(T)\right\Vert _{2}^{2} & ={\rm Tr}\left[\left(\hat{U}_{{\rm target}}^{\dagger}-\hat{U}^{\dagger}(T)\right)\left(\hat{U}_{{\rm target}}-\hat{U}(T)\right)\right]\\
 & ={\rm Tr}\left[\hat{U}_{{\rm target}}^{\dagger}\hat{U}_{{\rm target}}+\hat{U}^{\dagger}(T)\hat{U}(T)-\hat{U}_{{\rm target}}^{\dagger}\hat{U}(T)-\hat{U}^{\dagger}(T)\hat{U}_{{\rm target}}\right]\\
 & =2\left(d-{\rm Re}{\rm Tr}\hat{U}_{{\rm target}}^{\dagger}\hat{U}(T)\right)
\end{align*}
where $d$ is the underlying Hilbert space dimension. Thus, we see
that minimizing the error corresponds to maximizing the overlap ${\rm Re}{\rm Tr}\hat{U}_{{\rm target}}^{\dagger}\hat{U}(T)$
.

Now the real part looks suspicious -- if we have the gate right up
to a global phase, $\hat{U}(T)=e^{i\phi}\hat{U}_{{\rm target}}$ this
overlap indicates a non-perfect result. In fact, numerical experimentation
shows that this would be a serious drawback. We can trace this error
back to the original distance measure. The high-brow step to take
now would be to elevate the description to full quantum channels. Pragmatically,  we move from real part to absolute
square and thus the most common performance index for gates is
\[
J=\left|{\rm Tr}\left(\hat{U}_{{\rm target}}^{\dagger}\hat{U}(T)\right)\right|^{2}.
\]
This quantity can be interpreted in a somewhat operational fashion:
First apply the gate you have, then undo the gate you want. If everything
goes right you have but a global phase -- the same one on all vectors
of the standard basis. If not, you measure the deviation from unity
for the complete standard basis. There are other possible choices
(and good reasons to think about them), which we will discuss later.
With this quantity, we can proceed in a way similar to state transfer,
only that now we of course start at the unit matrix. We again use
piecewise constant controls and define both the intermediate propagator
and the intermediate back-propagated target
\[
\hat{X}_{j}=\hat{U}_{j}\cdots\hat{U}_{1}\quad\hat{P}_{j}=\hat{U}_{j+1}^{\dagger}\cdots\hat{U}_{N}^{\dagger}\hat{U}_{{\rm target}}
\]
 allowing us to rewrite $J=\left|{\rm Tr}\hat{P}_{j}^{\dagger}\hat{X}_{j}\right|^{2}$
for all values of $j$ . We can now apply the same identities as before
and find
\begin{align*}
\frac{\partial J}{\partial u_{i}(j)} & =\frac{\partial}{\partial u_{i}(j)}\left({\rm Tr}\hat{P}_{j}^{\dagger}\hat{X}_{j}\right)\left({\rm Tr}\hat{P}_{j}^{\dagger}\hat{X}_{j}\right)^{\ast}\\
 & =2{\rm Re}\left[\left(\frac{\partial}{\partial u_{i}(j)}{\rm Tr}\hat{P}_{j}^{\dagger}\hat{X}_{j}\right)\left({\rm Tr}\hat{P}_{j}^{\dagger}\hat{X}_{j}\right)\right]\\
 & =-2i\delta t{\rm Re}\left[\left({\rm Tr}\hat{P}_{j}^{\dagger}\hat{H}_{i}\hat{X}_{j}\right)\left({\rm Tr}\hat{P}_{j}^{\dagger}\hat{X}_{j}\right)\right].
\end{align*}
With this analytical gradient, the GRAPE algorithm can be applied
as above.

\subsection{The Krotov algorithms}

The \myiffindex{Krotov algorithm} \cite{Koch-Krotov-Main,Krotov1,krotov1983iteration,Krotov2}
has been formulated before the GRAPE algorithm. Some of its presentations
are historically based on applications in chemistry and emphasizes
constraints more than its core. Looking back on how GRAPE is applied,
we are blessed with an analytical gradient formula which in each iteration
allows us to calculate the gradient of the cost function(al) with
respect to all controls at all times and then by walking against it
look for improved controls. Notably, the gradient is always computed
at a point in parameter space given by the controls computed in the
\emph{previous }iteration.

There are two different algorithms which carry the name ``Krotov''
- a fact which can be quite confusing, even for experts in the field.

The first Krotov, prides itself with its monotonic convergence, which
is achieved by propagating the forward state using the old control
field, while the backward-propagating state makes use of the new field.
A detailed description, with Python implementation, can be found in
\cite{KrotovPython}.

The second Krotov can be considered a greedy version of GRAPE, and
is described in detail in \cite{Shai-PRA}: In this version of the
Krotov algorithm, all previously computed knowledge is used, i.e.,
once an entry to the gradient is computed, it is applied right away
and the next element of the gradient is computed with that correction
already applied. This approach of not leaving any information behind
in general lowers the number of iterations needed to reach convergence
and it comes with proven monotonic convergence. On the other hand,
each iteration step takes more time.

\begin{figure}
\includegraphics[width=0.9\columnwidth]{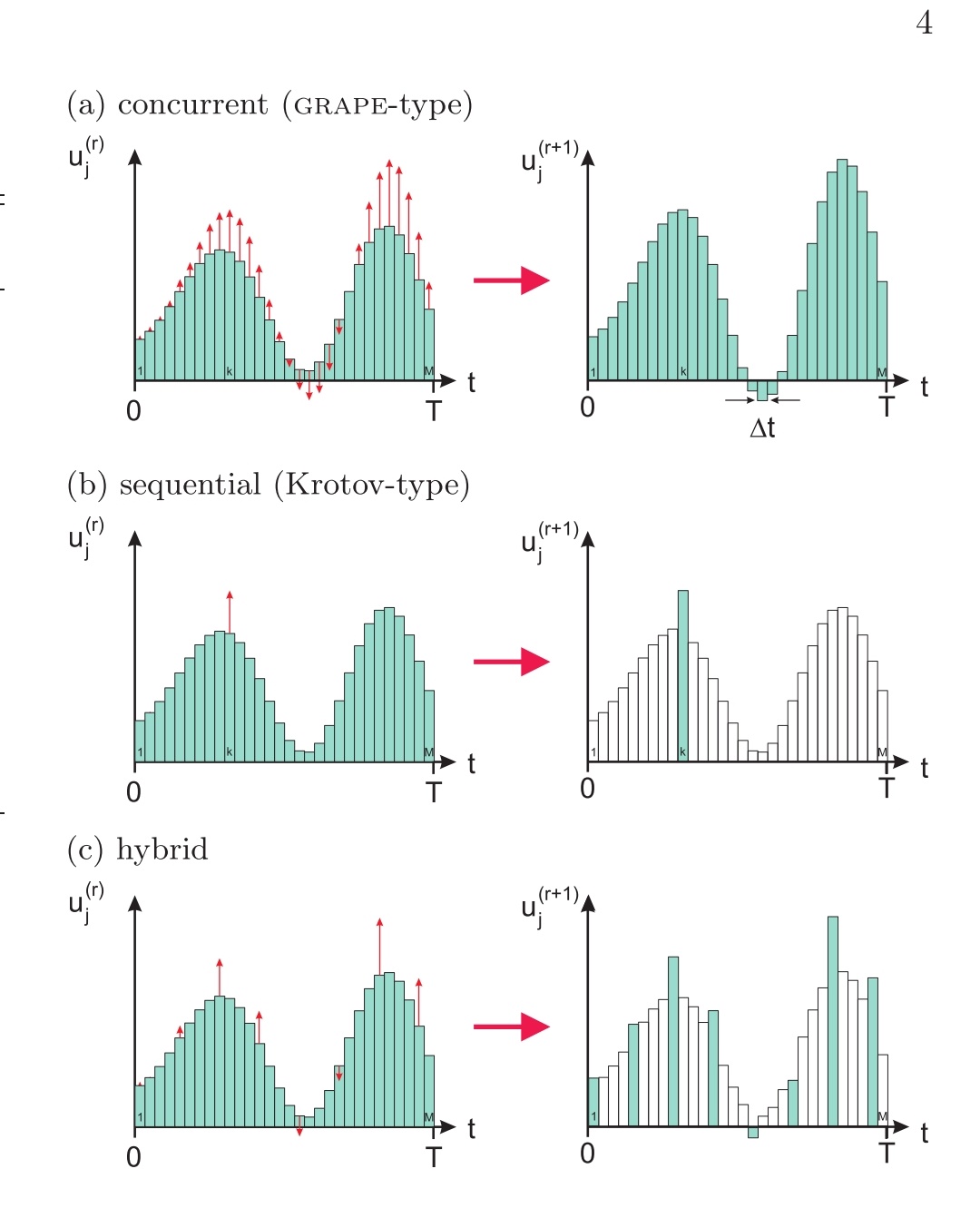}

\caption{Overview on the update schemes of gradient-based optimal control algorithms
in terms of the set of time slices T (q) = \{k(q), k(q), . . . k(q)
\} for which the control amplitudes are concurrently updated in each
iteration. Subspaces are enumerated by q, gradient-based steps within
each subspace by s, and r is the global step counter. In grape (a)
all the M piecewise constant control amplitudes are updated at every
step, so T (1) = \{1,2,...M\} for the single iteration q\ensuremath{\equiv}1.
Sequential update schemes (b) update a single time slice once, in
the degenerate inner-loop s\ensuremath{\equiv}1, be- fore moving to
the subsequent time slice in the outer loop, q; therefore here T (q)
= \{q mod M\}. Hybrid versions (c) follow the same lines: for instance,
they are devised such as to update a (sparse or block) subset of p
different time slices before moving to the next (disjoint) set of
time slices. \label{fig:update}}
\end{figure}

The various update strategies are visualized in figure \ref{fig:update}.

Benchmarking of the various optimal control algorithms is a topic
of ongoing research.

\subsection{Modern numerical issues}

\subsubsection{Control landscapes}

A gradient search with an analytical gradient as outlined is the best way to find a local extremum of an \myiffindex{optimization landscape}. If the optimization landscape has multiple local minima, it can get stuck in a local minimum and needs to be enhanced.

In a seminal series of papers, Rabitz has shown (see e.g. \cite{Rabitz2004}) that there is indeed only one extremum in the control landscape and that it is global. This theorem is a correct derivation of its assumptions -- one of which is the absence of constraints in pulse amplitude and temporal resolution. In practice, these constraints exist and multiple local extrema exist -- the more constrained the optimization, the more local extrema. Specifically in situations close to the quantum speed limit (see below), with low control resolution (Ref  \cite{Liebermann16} looks at a single bit of amplitude resolution and required genetic algorithms to converge) or with complex many-body dynamics and only few controls, these call for more advanced methods.

If one has a good intuition about the optimal pulse say, by solving a model that is very close to the desired model or by rescaling a solution that works at a longer gate duration, one can often stay close to the global extremum and otherwise requires a gradient search. If that is not the case, one needs to first start with a more global search method covering a large parameter space. Known systems for such gradient-free approaches are GROUP \cite{Sorensen18}, genetic algorithms (\cite{Judson1992,Liebermann16}), they are part of CRAB (see section \ref{ch:CRAB}) and simulated annealing \cite{Yang17}.

\subsubsection{Fidelities}

We would like to come back to the choice of \myiffindex{fidelity} based on the
2-norm described above. It has been argued that the most appropriate
way to characterize quantum processes is the use of the diamond norm \cite{quan-comp-accuracy1}. It can be expressed for a quantum operation $\mathcal{E}$ compared to an ideal operation $\mathcal{U}$ as
\begin{equation}
||\mathcal{U}_{\rm ideal}-\mathcal{E}||_{\diamond}=\sup_q \max_{\psi} \left|\rm{Tr}\left[\mathcal{U}_{\rm ideal}(|\psi\rangle\langle\psi|)-\mathcal{E}(|\psi\rangle\langle\psi|)\right]\right|
\end{equation}
This involves two generalizations of the 2-norm: On the one hand,
rather than taking the 2-norm distance which is equivalent to averaging
over all possible input states to the operation, we are taking the
maximum over $|\psi\rangle$, i.e., we choose the input state that maximizes the distance.
On the other hand, rather than directly using the unitary operation,
we enhance the Hilbert space by adding another space of dimension
$q$ on which the identity operation is performed. The diamond norm
is then the supremum over $q.$The latter may sound rather academic,
but it is not if, e.g., the initial state is entangled between the
original and the auxiliary system.

For the purposes of quantum optimal control, the diamond norm is rather
impractical -- it is hard to compute (as it contains a supremum) and
it can be non-differentiable (as it contains taking a maximum over
states, the state at which it has reached can jump in state space). What does this mean for
the applicability of quantum optimal control in the context of fault
tolerance?

There are two answers to this question. On the one hand, one can at
least find performance indices that emphasize the worst case more
strongly while being differentiable. A straightforward option is \cite{Hu08}
\begin{align*}
J_{q}&={\rm \max_{\alpha\in[0,2\pi)}}\left\Vert \hat{U}_{{\rm target}}-e^{i\alpha}\hat{U}(T)\right\Vert _{2q}^{2q}\\
&=\max_{\alpha\in[0,2\pi)}{\rm Tr}\left[\left(\hat{U}_{{\rm target}}^{\dagger}-e^{-i\alpha}\hat{U}^{\dagger}(T)\right)\left(\hat{U}_{{\rm target}}-e^{i\alpha}\hat{U}(T)\right)\right]^{q}
\end{align*}
which can be implemented in a straightforward fashion yet does not
have a known extension that avoids optimizing the global phase.

On the other hand, it is pragmatically not very crucial to go through
these steps as long as the algorithm converges properly: Our goal
is to get the error as close to zero as possible and, as these norms
can be continuously mapped onto each other, one pragmatically searches
for controls that reduce the error in the 2-norm to an extremely low
value which guarantees that even in the desired norm the error is
low enough -- using the paradigm to control and verify with two different
measures.

\subsubsection{Increasing precision of GRAPE}

The GRAPE algorithm above defines a straightforward gradient algorithm
for optimal control. There are a few known measures to speed up its
convergence.

One measure is the improvement of the use of the gradient by moving
to a quasi-Newton method, the Broyden, Fletcher, Goldfard, and Shanno
(BFGS) method \cite{L-BFGS}. Newton's method, as the reader may have
learned in an elementary introduction to numerical mathematics, rely
on approximating the function whose zero we desire to find by its
tangent -- in our case, we desire to find the zero of the gradient,
i.e., we need to approximate the functions up to its second derivative.
As we are optimizing a scalar that depends on many parameters -- all
the controls taken at all the times of interest -- the matrix of second
derivatives is a high-dimensional object. In order to approximate
the zero of the gradient, one would have to invert that matrix,
which is numerically hard and would likely negate the potential computational
advantage. The BFGS method instead relies on directly approximating
the inverse Hessian

\subsubsection{\label{subsec:The-gradient-of-a-matrix-exponential}The gradient
of a matrix exponential}

Expanding on the discussion surrounding eq. (\ref{eq:PWC-Uk}), (\ref{eq:Deriv-of-time-slice}),
any gradient-driven optimal control optimization, such as GRAPE or
Krotov, which treats the control fields as piecewise constant, will
describe the coherent propagator of time slice $m$ as
\begin{equation}
U_{m}=\exp\left(-\frac{i}{\hbar}\delta_{t}H\left(\bar{\alpha},t_{m}\right)\right)\label{eq:PWC-U-m}
\end{equation}

where $\bar{\alpha}$ parameterizes the control functions $u\left(t\right)$.
We are searching for the value of $\bar{\alpha}$ which will minimize
the infidelity. At step $j$ of the optimization, to compute the gradient
of the goal function with respect to $\bar{\alpha}$, we must compute
$\partial_{\bar{\alpha}}U_{m}\left(\bar{\alpha}\right)\vert_{\bar{\alpha}=\bar{\alpha}_{j}}$
. At this point we can rewrite eq. (\ref{eq:PWC-U-m}) as in eq. (\ref{eq:Deriv-of-time-slice}),
\[
U_{m}=\exp\left(-\frac{i}{\hbar}\delta_{t}\left(H_{m,j}+\epsilon_{\bar{\alpha}}\tilde{H}_{m,j}\right)\right)
\]
where $\epsilon_{\bar{\alpha}}$ is small and we seek $\partial_{\epsilon_{\bar{\alpha}}}U_{m}\left(\epsilon_{\bar{\alpha}}\right)\vert_{\epsilon_{\bar{\alpha}}=\bar{0}}$.
Following \cite{grad-exp-1-aizu1963parameter,exp-grad-2-levante1996pulse},
and their summary in Appendix A of \cite{Shai-PRA}, we denote
the eigenvalues and eigenvectors of $H_{j}$ by $e_{k}$ and $\ket{e_{k}}$,
respectively, then using the spectral theorem
\[
\Braket{e_{l}}{\partial_{\bar{\epsilon_{\bar{\alpha}}}}U_{m}}{e_{k}}=\left\{ \begin{array}{ccc}
-\frac{i}{\hbar}\delta_{t}\Braket{e_{l}}{\tilde{H}_{m,j}}{e_{k}}\exp\left(-\frac{i}{\hbar}\delta_{t}e_{l}\right) &  & \textrm{if\,\,}e_{l}=e_{k}\\
-\frac{i}{\hbar}\delta_{t}\Braket{e_{l}}{\tilde{H}_{m,j}}{e_{k}}\dfrac{\exp\left(-\frac{i}{\hbar}\delta_{t}e_{l}\right)-\exp\left(-\frac{i}{\hbar}\delta_{t}e_{k}\right)}{-\frac{i}{\hbar}\delta_{t}\left(e_{l}-e_{m}\right)} &  & \textrm{if\,\,}e_{l}\neq e_{k}
\end{array}\right.
\]

one may invoke the spectral theorem in a standard way and calculate
matrix functions via the eigendecomposition.

To simplify notation, we shall look at $\partial_{x}\;e^{A+xB}$,
with $A,B$ being an arbitrary pair of Hermitian (non-commuting) matrices
and $x\in\mathbb{R}$. As previously $\{\ket{e_{l}}\}$ as the orthonormal
eigenvectors to the eigenvalues $\{e_{l}\}$ of $A$
. We then obtain the following straightforward, if somewhat lengthy,
derivation:

\begin{eqnarray*}
D & = & \braket{e_{l}}{\partial_{x}\;e^{A+xB}|e_{k}}\Big|_{x=0}\\
 &  & \braket{e_{l}}{\partial_{x}\;\sum_{n=0}^{\infty}\frac{1}{n!}\big(A+xB\big)^{n}|e_{k}}\Big|_{x=0}\\
 &  & \braket{e_{l}}{\sum_{n=0}^{\infty}\frac{1}{n!}\sum_{q=1}^{n}\big(A+xB\big)^{q-1}B\big(A+xB\big)^{n-q}|e_{k}}\Big|_{x=0}\\
 &  & \braket{e_{l}}{\sum_{n=0}^{\infty}\frac{1}{n!}\sum_{q=1}^{n}A^{q-1}BA^{n-q}|e_{k}}\\
 &  & {\sum_{n=0}^{\infty}\frac{1}{n!}\sum_{q=1}^{n}e_{l}^{q-1}\braket{e_{l}}{B|e_{k}}e_{k}^{n-q}}\\
 &  & \braket{e_{l}}{B|e_{k}}\sum_{n=0}^{\infty}\frac{1}{n!}\sum_{q=1}^{n}e_{l}^{q-1}e_{k}^{n-q}
\end{eqnarray*}

This provides the answer for in the case where $e_{l}=e_{k}$. For
$e_{l}\neq e_{k}$ a bit more work is needed:

\begin{eqnarray*}
D & = & \braket{e_{l}}{B|e_{k}}\sum_{n=0}^{\infty}\frac{1}{n!}e_{k}^{n-1}\sum_{q=1}^{n}\left(\frac{e_{l}}{e_{k}}\right)^{q-1}\\
 &  & \braket{e_{l}}{B|e_{k}}\sum_{n=0}^{\infty}\frac{1}{n!}e_{k}^{n-1}\frac{(e_{l}/e_{k})^{n}-1}{(e_{l}/e_{k})-1}\\
 &  & \braket{e_{l}}{B|e_{k}}\sum_{n=0}^{\infty}\frac{1}{n!}\frac{e_{l}^{n}-e_{k}^{n}}{e_{l}-e_{k}}\\
 &  & \braket{e_{l}}{B|e_{k}}\frac{e^{e_{l}}-e^{e_{k}}}{e_{l}-e_{k}}
\end{eqnarray*}

Note that we have explicitly made use of the orthogonality of eigenvectors
to different eigenvalues in normal matrices.

\section{Applied optimal quantum control}

\begin{figure}[ht]\centering
\includegraphics{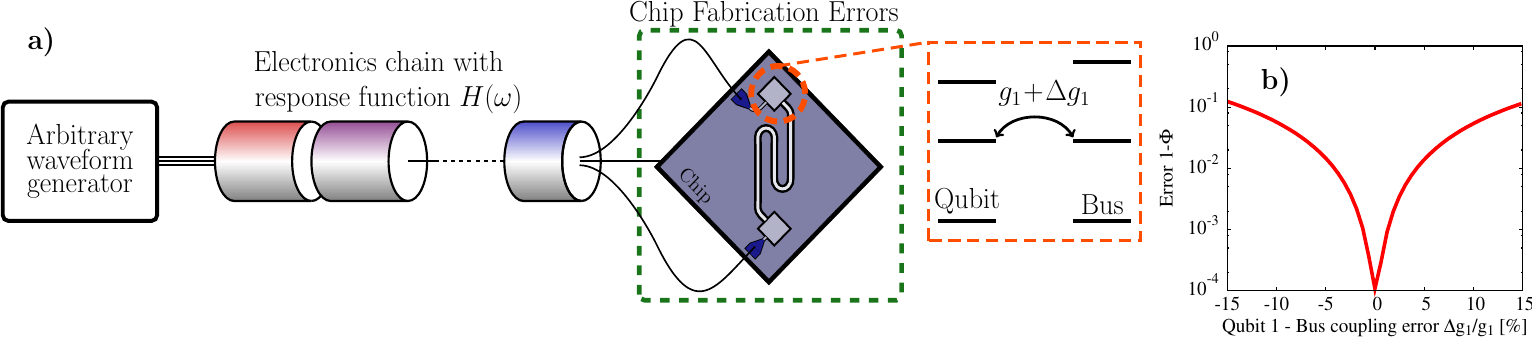}
\caption{\label{fig:chain} Typical sources of inaccuracy in quantum control for superconducting qubits including the transmission from the generator to the sample and inaccuracy of the Hamiltonian model. Right: Typical error sensitivity for a gate between superconducting qubits.  }
\end{figure}

While quantum optimal control is a well-developed field and has been
very successful in atomic and molecular systems, its track record
in solid-state quantum technologies is somewhat less developed.
The reason has to do with the accuracy of the models, i.e., the precision
at which we know every ingredient of the Hamiltonian. First of all,
a quantum-technological device (specifically, but not exclusively,
in the solid state) has human-made components which contain some fabrication
uncertainty. This affects the drift Hamiltonian -- even if its eigenvalues
can be accurately determined using spectroscopy, it is much more involved
to find its eigenvectors. These naturally also affect the matrix elements
of the control Hamiltonians. On top of that, some solid-state quantum
devices need to be extremely well isolated from their environments
including high-temperature black-body radiation. This means, that an
applied control signal will get distorted on its way to the sample
in a way that can be measured only to a limited degree, see fig. \ref{fig:chain} for a summary. While one can
improve hardware and characterization to meet these challenges, it
is hard to get this to the precision required by, say, fault-tolerant
quantum computing. Thus, other approaches are called for.

\subsection{Closing the loop for pulse calibration}

One possible approach to handle uncertainties would be to use a robust
control methodology inspired by magnetic resonance in ensembles. While
this method can be useful, it slightly misses the point: It still
requires a good estimate for the uncertainty and then it improves
performance across the relevant parameter interval. Here, the situation
is different, we do not have a parameter distribution but a single
set of parameters -- we just cannot find it or even the relevant model
\emph{a priori.}

One way to still find good pulses are hybrid control methods such
as Adaptive Hybrid Optimal Control (AdHOC, \cite{AdHOC}), Optimized Randomized Benchmarking for Immediate Tuneup (ORBIT, \cite{ORBIT}), and Adaptive Control via Randomized Optimization Nearly Yielding Maximization \phantom{spazzinate!} (ACRONYM, \cite{Ferrie15}). The idea of these methods is rather similar:
After an initial design phase that may or may not contain traditional
optimal control, a set of pulses is constructed based on models that
are believed to approximate the actual system but its
parameterization is left open to some corrections. These corrections
are then determined in a closed loop -- the fidelity is
measured and the pulses are updated based on these fidelity measurements.

\begin{figure}[ht]\centering
\includegraphics{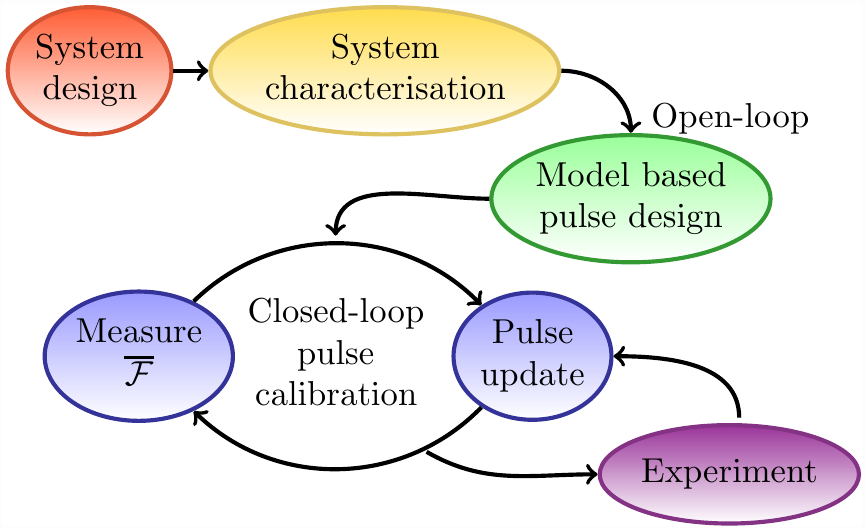}
\caption{\label{fig:adhoc}  Typical two-stage control workflow with an open loop modeling stage}
\end{figure}

In the example of AdHOC, the pulse measurement is based on randomized
benchmarking (described below) and the optimization that determines
the corrections is based on the \myiffindex{Nelder-Mead} simplex algorithm, which
is available in most numerical mathematics toolboxes. What is crucial is that this is a gradient-free algorithm in order to avoid issues with taking gradients of measurement data.
Is that as a simplex algorithm, the search for a pulse described by
$n$ parameters needs to be initialized using $n+1$ initial guesses.
This raises the important question how the number $n$ can be kept
as small as possible (but not smaller, see below) by finding an efficient
parameterization.

This is not an easy problem. So far, we have always assumed that the
pulses are parameterized in piecewise constant fashion and have argued
that this is naturally compatible with arbitrary wave form generators.
However, this parameterization does not naturally lend itself to reduction
of the number of parameters -- simple, sparse controls in quantum physics
are typically sine and cosine functions with smooth, Gaussian-derived
envelopes. On the other hand, the piecewise constant parameterization
was instrumental in deriving the gradient formula in an analytical
way and cannot be easily removed.

\subsection{CRAB\label{ch:CRAB}}

Albeit originally developed from a different motivation, the optimization
of many-body dynamics, the \myiffindex{Chopped RAndom Basis (CRAB)}\footnote{pronounced with a rolling 'r' and a voiced 'b'}
algorithms serves that purpose, \cite{CRAB}. It introduced the concept
of simple and sparse pulse parameterizations, i.e., finding a pulse parameterization that is not necessarily piecewise constant but rather can be written as
\begin{equation}
H\left(\bar{\alpha},t\right)=H_{0}+\sum_{k=1}^{C}c_{k}\left(\bar{\alpha},t\right)H_{k}\,,
\label{eq:Ham_tot}
\end{equation}
where the functions $c_k$ can e.g. be harmonic functions characterized by amplitude, frequency and phase or a sequence of Gaussians
\begin{equation}
c_{k}\left(\bar{\alpha},t\right)=\sum_{j=1}^{m}A_{k,j}\exp\left(-(t-\tau_{k,j})^2/\sigma_{k,j}^2\right).
\label{eq:Fourier-controls}
\end{equation}
In  complex systems that were the initial motivation for CRAB, one has very little prior knowledge about a suitable basis and it is at best chosen random, hence the name. CRAB utilizes a gradient-free
search, specifically Nelder-Mead (although other algorithms could
be used), similarly to what we have already described for AdHOC\footnote{note that CRAB was proposed before AdHOC}
.

The fact CRAB is model-free, with the gradient-free search treating
the quantity to be optimized as a black box, provides a distinct advantage
in situations a precise model is unknown or when the model is know,
but the gradient cannot be computed due to numerical complexity or
other reasons. This makes CRAB appropriate for closed-loop experimental
calibration of control fields in system ranging from nitrogen vacancy
centers in nano-diamonds \cite{CRAB-Appl-1-binder2017qudi} and cancer
treatment formulations \cite{CRAB-Appl-2-angaroni2019personalized},
to DMRG-based simulations \cite{CRAB-Appl-3-silvi2014lattice}. Further,
CRAB enjoys huge success in studying quantum phase transitions,
preparing large Schrödinger cat states, sensing and many more.

A variant of CRAB, known as dCRAB \cite{dCRAB}, deals with a situation
where the control parameterization has a higher dimensionality than
can be optimized by Nelder-Mead, by iteratively optimizing different
subsets (or low-dimension projections) of the high-dimension full
parameter space.

\subsection{GOAT}

\myiffindex{Gradient Optimization of Analytic conTrols (GOAT)} is a recently \cite{Machnes18} proposed
optimal control algorithm which does not derive from the variational
formulation of optimal control, defined earlier. Rather, GOAT finds
the equations of motion for the gradient of the goal function with
respect to the control parameters, integrating as you would the Schrödinger
equation (as piece-wise-constant approximation, or using standard
ODE tools such as Runge-Kutta optimizers).

For our purpose, the goal function to minimize is defined as the projective
$SU$ distance (infidelity) between the desired gate, $U_{\textrm{goal}}$,
and the implemented gate, $U\left(T\right)$, \cite{PalaoKosloff2002}
(also \cite{TESCH2001633})
\begin{equation}
g\left(\bar{\alpha}\right):=1-\tfrac{1}{\textrm{dim}\left(U\right)}\left|\textrm{Tr}\left(U_{\textrm{goal}}^{\dagger}U\left(T\right)\right)\right|\,,\label{eq:goal}
\end{equation}
where $U\left(t\right)$ is the time ordered ($\mathbb{T}$) evolution
operator
\begin{equation}
U\left(\bar{\alpha},T\right)=\mathbb{T}\exp\left(\int_{0}^{T}-\frac{i}{\hbar}H\left(\bar{\alpha},t\right)dt\right).\label{eq:Upt}
\end{equation}

GOAT's ability to use any control ansatz makes it feasible to find
drive shapes described by a small number of parameters, suitable for
closed-loop calibration.

A gradient-based optimal control algorithm requires two ingredients:
an efficient computation of $\partial_{\bar{\alpha}}g\left(\bar{\alpha}\right)$
and a gradient-based search method over parameter space. GOAT presents
a novel method for the former, while using any standard algorithm
for the latter, such as BFGS.

Consider the gradient of the goal function eq. (\ref{eq:goal}) with
respect to $\bar{\alpha}$,
\begin{equation}
\partial_{\bar{\alpha}}g\left(\bar{\alpha}\right)=-\textrm{Re}\left(\frac{g^{*}}{\left|g\right|}\frac{1}{\textrm{dim}\left(U\right)}\textrm{Tr}\left(U_{\textrm{goal}}^{\dagger}\partial_{\bar{\alpha}}U\left(\bar{\alpha},T\right)\right)\right)\,.\label{eq:goal-func-grad}
\end{equation}
Neither $U\left(\bar{\alpha},T\right)$ nor $\partial_{\bar{\alpha}}U\left(\bar{\alpha},T\right)$
can be described by closed form expressions. $U$ evolves under the
equation of motion $\partial_{t}U\left(\bar{\alpha},t\right)=-\frac{i}{\hbar}H\left(\bar{\alpha},t\right)U\left(\bar{\alpha},t\right)$.
By taking the derivative of the $U$ equation of motion with respect to $\bar{\alpha}$
and swapping derivation order, we arrive at a coupled system of equations of motion
for the propagator and its gradient,
\begin{equation}
\partial_{t}\left(\begin{array}{c}
U\\
\partial_{\bar{\alpha}}U
\end{array}\right)=-\frac{i}{\hbar}\left(\begin{array}{cc}
H & 0\\
\partial_{\bar{\alpha}}H & H
\end{array}\right)\left(\begin{array}{c}
U\\
\partial_{\bar{\alpha}}U
\end{array}\right).\label{eq:joint-eom}
\end{equation}
As $\bar{\alpha}$ is a vector, $\partial_{\bar{\alpha}}U$ represents
multiple equations of motion, one for each component of $\bar{\alpha}$.
$\partial_{\bar{\alpha}}H$ is computed using the chain rule.

GOAT optimization proceeds as follows: Starting at some initial $\bar{\alpha}$
(random or educated guess), initiate a gradient driven search (e.g.
L-BFGS \cite{L-BFGS}) to minimize eq. (\ref{eq:goal}). The search
algorithm iterates, requesting evaluation of eqs. (\ref{eq:goal},\ref{eq:goal-func-grad})
at various values of $\bar{\alpha}$, and will terminate when the
requested infidelity is reached or it fails to improve $g$ further.
Evaluation of $g\left(\bar{\alpha}\right)$, $\partial_{\bar{\alpha}}g\left(\bar{\alpha}\right)$
requires the values of $U\left(\bar{\alpha},T\right)$ and $\partial_{\bar{\alpha}}U\left(\bar{\alpha},T\right)$.
These are computed by numerical forward integration of eq. (\ref{eq:joint-eom}),
by any mechanism for integration of ordinary differential equations that is accurate and efficient
for time-dependent Hamiltonians, such as adaptive Runge-Kutta. Initial
conditions are $U\left(t=0\right)=\mathcal{I}$ and $\partial_{\bar{\alpha}}U\left(t=0\right)=0$.
Note that no back propagation is required.

Experimental constraints can be easily accommodated in GOAT by mapping
the optimization from an unconstrained space to a constrained subspace,
and computing the gradient of the goal function using the chain rule.
For example, $\bar{\alpha}$ components may be constrained by applying
bounding functions, e.g. $\alpha_{k}\longrightarrow\frac{1}{2}\left(v_{\text{max}}-v_{\text{min}}\right)\sin\left(\bar{\alpha}_{k}\right)+\frac{1}{2}\left(v_{\text{max}}+v_{\text{min}}\right)$
which imposes $\alpha_{k}\in\left[v_{\text{min}}\ldots v_{\text{max}}\right]$.
Amplitude constraints and a smooth start and finish of the control
pulse can be enforced by passing the controls through a window function
which constrains them to a time-dependent envelope. Gradients for
$\partial_{\bar{\alpha}}H$ flow via the chain rule.

\subsection{Evaluating fidelity with randomized benchmarking.}

The closed-loop approaches mentioned above crucially rely on a measurement
of success. While in state-transfer problems, e.g. creating an ordered
state quickly or steering a chemical reaction, there may be generic
tools to determine this success with a given experimental ap\-par\-atus.
In the case of a quantum gate, this is not so simple. While classic
textbooks like  first label quantum \myiffindex{process tomography}, this has
a number of drawbacks, and is now replaced by more efficient methods.

\subsubsection{The trouble with tomography}

To understand this, let's first take a look at quantum state tomography \cite{NielsenChuang}.
This is, in a nutshell, the reconstruction of a quantum state (characterized
by its density matrix) by performing a complete set of observable
measurements. Next to some practical drawbacks having to do with guaranteeing
a positive density matrix \cite{Paris04}, this is also impractical: A typical quantum
device can be read out with a single machine -- an electric or optical
measurement. Formally this corresponds to measuring in one basis (we
will assume that we are dealing with qubits, so recording the expectation
value completely characterizes the output distribution). In order
to measure a complete set of operators, one has to first perform a
basis change in the shape of performing a coherent operation. As this
operation itself is prone to error, this will falsify the result.
Together with the intrinsic imperfection of the readout device this
constitutes measurement error.

From state tomography, it is another step to process tomography, i.e.,
the reconstruction of a quantum channel -- linear map from input to
output density matrices -- from measured. Formally, one can using
the Choi-Jamiolkowski isomorphism \cite{NielsenChuang} map the process matrix of the channel
onto the density matrix of a state and treat the problem of process
tomography as one of state tomography. Practically, process tomography
involves to now measure complete sets of both initial and final states
that undergo the channel. Similar to measurement, also state preparation
is usually possible only in one distinct basis -- if state preparation
is performed by measurement it is the measurement basis, if state
preparation is performed via thermalization or optical pumping it
is the drift Hamiltonian's eigenbasis -- and it is imperfect -- both
of these give rise to state preparation errors. Thus, in total, the
quantum channel that one would like to characterize is masked by state
preparation and measurement (SPAM) errors.

\begin{figure}[ht]\centering
\includegraphics[width=\textwidth, clip, trim={0 6cm 0 15cm}]{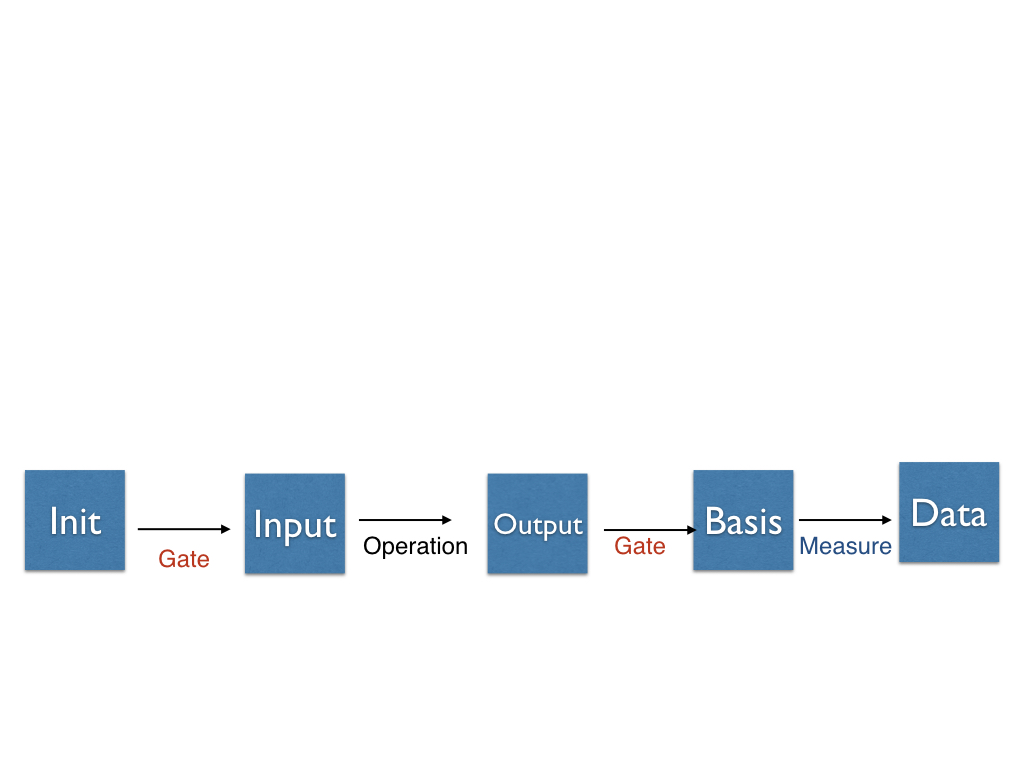}
\caption{\label{fig:spam}  SPAM errors in process tomography: The state that can be prepared and measured needs to be transferred into the basis that has to be prepared and measured, introducing additional errors obfuscating the channel.}
\end{figure}

On top of that, full process tomography is also forbiddingly labourious.
The state of a $d$ dimensional quantum system is characterized by
a $d^{2}$ entries in a densitry matrix that, accounting for hermiticity
and norm boil down to $d^{2}-d+1$ real numbers. This has to be squared
again to describe a quantum channel, leading to $O\left(d^{4}\right)$
numbers -- which then are recombined to compute a single fidelity.
In an $n$ - Qubit system, we have $d=2^{n}$ making full tomography
forbiddingly data intensive. On top of that, we would like to ensure
complete positivity of the measured channel, which gives rise to inequality
constraints that are practically hard to meet specifically when the
map is close to unitary. Now there are several methods such as compressed
sensing and Monte Carlo sampling \cite{Gross10,Chasseur17} that reduce that problem, but with
SPAM still included, there is strong motivation to look for an independent
method to evaluate fidelity in an experiment. Here, randomized benchmarking
and its descendants (RB+) have appeared as a quasi-standard. A comprehensive review of RB+ has currently not been published. We are going to mention key papers on the way and otherwise refer to the work of J. Emerson.

\subsubsection{Randomization of quantum channels}

Let's first lay the foundation of how we describe a \myiffindex{quantum channel} \cite{NielsenChuang}:
A linear map hat takes any valid density matrix onto another valid
density matrix, i.e., with
\[
\rho\mapsto\mathcal{E}\left[\rho\right]
\]
 we demand that if $\rho$ is hermitian, positively semidefinite,
and has a normalized trace, so is $\mathcal{E}\left[\rho\right]$.
This is satisfied by the iffindex{Kraus representation}
\[
\mathcal{E}\left[\rho\right]=\sum_{k}\hat{A}_{k}\hat{\rho}\hat{A}_{k}^{\dagger}\quad\sum_{k}\hat{A}_{k}^{\dagger}\hat{A}_{k}=\mathbb{I}.
\]
 The (non-unique) \emph{Kraus operators }$A_{k}$ characterize the
channel. It can be easily shown that the Kraus representation leads
to a valid channel and it takes a bit more attention to show that
the validity of the channel also requires the Kraus representation.

Now to estimate the average fidelity over a channel  relative to
a desired unitary $\hat{U}$ we apply the channel to a pure initial
state, then undo the ideal channel, compute the overlap with the pure
state and average over all pure inputs
\[
F=\int d\psi\ \left\langle \psi\left|U^{\dagger}\mathcal{E}\left[\left|\psi\rangle\langle\psi\right|U\right]\right|\psi\right\rangle
\]
 where the integral runs over a suitable uniform distribution of all
states called the Haar measure. We now aim at replacing the average in this formula by another randomization procedure \cite{Emerson_2005}. We now decompose the real operation
into an ideal operation followed by an error channel and Kraus-decompose
the error channel
\[
\mathcal{E}=\Lambda\circ\mathcal{U}\quad\Lambda=\sum_{k}A_{k}\rho A_{k}^{\dagger}.
\]
 Plugging this into the expression for the average gets us
\[
F=\int d\psi\ \left\langle \psi\left|U^{\dagger}\Lambda\left[U\left|\psi\rangle\langle\psi\right|U^{\dagger}\right]U\right|\psi\right\rangle .
\]
 We can read this expression as implementing the motion-reversal transformation
$U^{\dagger}\cdot U$ with an error $\Lambda$ occuring in the middle.

Now instead of going for $F$ directly, let us average the fidelity
over all unitaries that can enter the motion-reversal map -- assuming
tacitly that we have the same $\Lambda$ at all times. We now
compute a at first glance very different average -- we keep a single
initial state $\rho=|\psi\rangle\langle\psi|$ and instead average
over all unitaries
\[
E=\int dU\ {\rm Tr}\left[\rho U^{\dagger}\Lambda\left(U\rho U^{\dagger}\right)U\right].
\]
Now we exchange the order of integration and change the order under
the trace and write this as
\[
E={\rm Tr}\left(\rho\left[\int dU\ U\Lambda U^{\dagger}\right]\rho\right)
\]

We can now read this exchanged expression at face value -- in the center
is noise averaged over all unitaries
\[
\Lambda_{{\rm ave}}=\int dU\ U\Lambda U^{\dagger}.
\]
Building on the operations of unitary maps as generalized rotations,
this is called a twirled channel. It can be mathematically shown what
is physically rather obvious -- this channel must be highly symmetric,
it cannot prefer any basis over the other. The only channel compatible
with this is the depolarizing channel
\[
\Lambda_{{\rm ave}}\left[\rho\right]=p\rho+\frac{1-p}{d}\mathbb{I}
\]
which has a single error probability $p.$ With this the error averaged
over all unitaries equals the fidelity of the twirled channel computed
for a single input state
\[
E={\rm Tr}\left(\rho\Lambda_{{\rm ave}}\rho\right)=F.
\]
where the last equality requires some more involved math to show that
this is also the same as the average fidelity of a unitary averaged
over all states. The fact that a single input state is enough -- we
have delegated the need for averaging from all states to twirling
the channel -- addresses the problem of SPAM errors.

Now what is needed is an efficient way to implement $\Lambda_{{\rm ave}}$.
We need to replace the integral over all unitaries by a sum over random
elements that converges to this integral. This brings in the concept
of a unitary 2-design: a set that correctly reproduces the full unitary
set in polynomials of degree 2 . It can be shown \cite{Dankert09,Knill09} (in a rather pedestrian
way) that the Clifford group is sufficient. The Clifford group \cite{NielsenChuang} is
formally defined as the normalizer of the Pauli group. For n qubits,
this Pauli $P_{n}=\left\{ \text{\ensuremath{\sigma_{n}}}\right\} $
group consists of all direct products of Pauli matrices $\sigma_{n}=\otimes_{j=1}^{n}\sigma_{i_{j}}$,
$i_{j}\in\left\{ 0,1,2,3\right\} $ so the corresponding Clifford
group is the set of all unitaries that map all n-qubit Pauli matrices
onto Pauli matrices
\[
C_{n}=\left\{ U\in SU\left(2^{n}\right):\forall\sigma_{n}\in P_{n}\ \exists\sigma_{m}\text{\ensuremath{\in}}P_{m}:\sigma_{m}=U\sigma_{n}U^{\dagger}\right\} .
\]
 For a single qubit, this group is generatd by all quarter-turns around
the Bloch sphere. The Clifford group is a discrete group and quantum
algorithms consisting of only Clifford gates can be efficiently classically
simulated . These together lead to the remarkably simple protocol
of randomized benchmarking.

\subsubsection{Randomized Benchmarking}

Let's pull all of these ingredients together into a handy protocol:
\begin{enumerate}
\item Repeat for a few representative sequences
\begin{enumerate}
\item Draw a random set of Clifford gates
\item Compute the resulting operation and its inverse. Add the inverse to
the end of the sequence
\item Repeat the following to establish an estimate for the final probability
for survival of the initial state
\begin{enumerate}
\item initialize the system in a convenient state
\item run the sequence
\item measure if the outcome is the same state or not
\end{enumerate}
\item Average to estimate the survival probability for the given sequence
\end{enumerate}
\item Average to estimate the survival probability averaged of the Clifford
groups. As a function of sequence length, the result will have the
form
\[
p(n)=p_{0}+\lambda^{n}.
\]
Here, $\lambda$ is the average Clifford gate fidelity and can be
determined by fitting, whereas $p_0$ is the SPAM error.
\end{enumerate}
It turns out practically and can be reasoned analytically that the
need for averaging is acceptable, artifacts of ensemble sizes vanish
quickly \cite{Chasseur15}.

In this basic version of RB, there are a lot of assumptions that can
be questioned. The theory of randomized benchmarking has been extended
to adapt most of the demands resulting from weakening these assumptions.
We cannot do the vast literature full justice here but mention a few
highlights.

First of all, standard RB finds the fidelity averaged over the whole
Clifford group. If one instead desires to characterize a single Clifford
gate, the technique of interleaved randomized benchmarking (IRB) \cite{Magesan12} can
be applied. There, one first performs regular RB. Then, one takes
the sequences used for RB and interleaves the desired Clifford gate
between any two of the gates from the sequence. The inverse to the
resulting sequence needs to be re-computed. The comparison between
the interleaved and the regular frequencies gives the average fidelity
of that special Clifford gate.

\begin{figure}[ht]\centering
\includegraphics[width=\textwidth, clip, trim={0 15cm 0 5cm}]{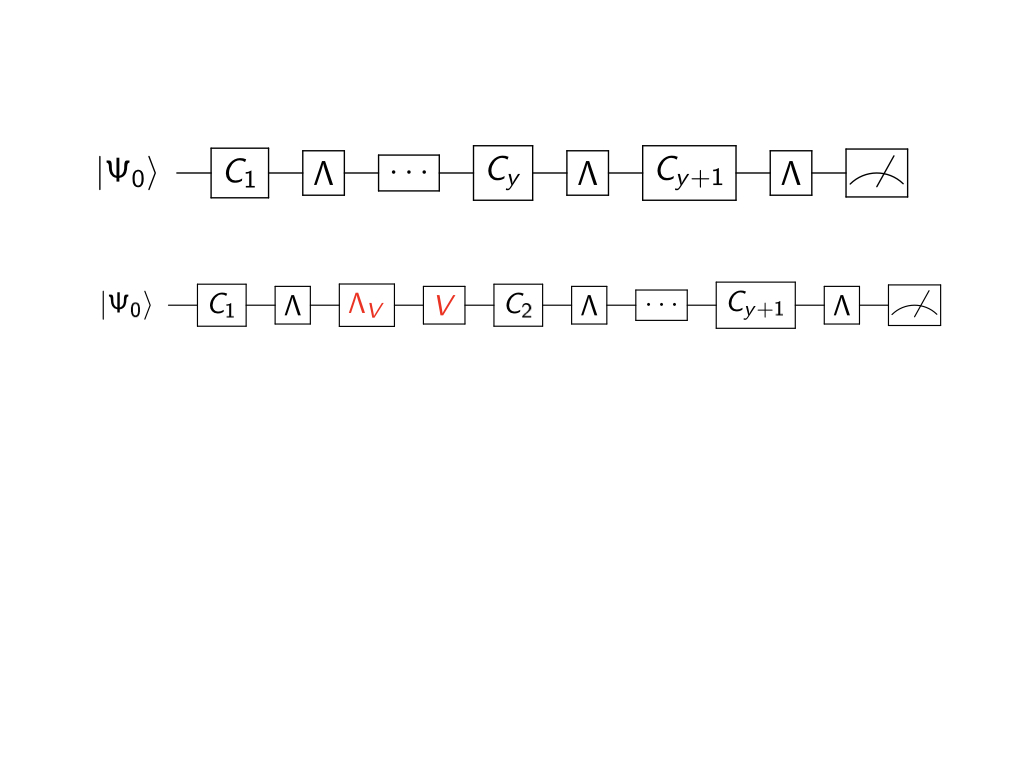}
\caption{\label{fig:rbirb}  Quantum channels for randomized benchmarking. Top: Randomized Benchmarking consists of a sequence of random (perfect) Clifford $C_i$ and Errors $\Lambda$ inverted by the last Clifford gate $C_{y+1}$. Bottom: Interleaved randomized Benchmarking interleaves a particular Clifford gate $V$ into this sequence. }
\end{figure}

In a similar vein, issues like leakage out of the computational subspace,
gate-dependent error and others can be taken into account \cite{Chasseur15}, leading
to the modern concept of cycle benchmarking. Including non-Clifford
gates, however, can only be done at the cost of significant overhead,
as the inverting operation is hard to compute as well as hard to invert
- it is an arbitrarily quantum gate encompassing the whole system
and not part of the Clifford group. A combination of RB with Monte
Carlo sampling can be applied to still keep parts of the benefits
of IRB \cite{Chasseur17}.

That being said, in many practical architectures, the only non-Clifford
gate is the T-gate, a $\pi/4$ z-axis rotation which can be done
in software to high precision, so it is not crucial to calibrate it
with optimal control. Also, as the two-qubit CNOT gate is a Clifford
gate, one cannot claim that natively and without error correction
Cliford gates are easier than non-Clifford.

\subsection{Approximating time evolutions with the Magnus expansion \label{ch:Magnus}}

Control calculations involve solving the time-dependent Schrödinger
equation. While this can be done analytically in, e.g., rotating wave
situations or approximations, this can quickly become hard -- even
for a system as simple as a harmonically driven two-state-system
this is a daunting task \cite{Gangopadhyay2010}. If we would like to proceed analytically with optimal control as far as possible, computing the final gate analytically is a key ingredient to which the Magnus expansion is an important ingredient.
Numerically, techniques for coupled ordinary
differential equations like Runge-Kutta can be used as well as split-operator
techniques. For analytical calculations, one can use the Dyson series
familiar from regular advanced quantum mechanics  as systematic perturbation
theory. In many cases, it is however more effective to use the 
\myiffindex{Magnus expansion}, an asymptotic expansion that used the number of nested
commutators as a small parameter. It is exact  but usually truncated
at low order. Our treatment mostly follows \cite{WarrenWarren}.

The problem at hand is to start from a Hamiltonian that has a (hopefully)
large but solvable component and a perturbation $\hat{H}=\hat{H}_{0}(t)+\hat{V}(t)$.
A clever choice of this division is key and there is no need for the
former to be time-independent. We can transfer to the interaction
picture with respect to $\hat{H}_{0}$ . The resulting transformed
perturbation $\hat{V}^{I}(t)$ will then acquire additional time-dependence,
often in the form of large oscillating terms. The objective is now
to approximately calculate the time evolution
\[
\hat{U}^{I}(t)=\mathbb{T}\exp\left(-\frac{i}{\hbar}\int_{0}^{t}d\tau\ \hat{V}^{I}\left(\tau\right)\right)
\]
 where $\mathbb{T}$ is the usual time ordering operator. The Dyson
expansion of this term starts as
\[
\hat{U}^{I}\left(t\right)=1-\frac{i}{\hbar}\int_{0}^{t}d\tau\ \hat{V}^{I}\left(\tau\right)-\frac{1}{\hbar^{^{2}}}\int_{0}^{t}d\tau\ \int_{0}^{\tau}d\tau^{\prime}\ \hat{V}^{I}(\tau)\hat{V}^{I}\left(\tau^{\prime}\right)+\dots
\]
 which we can expect to converge quickly if the perturbation combined
with oscillations are so small that the integration over (potentially)
long times does not hinder convergence. If this is not the case, one
could resort to self-energy techniques as they are known in quantum
field theory. For these time-dependent systems, the Magnus expansion
is a related route. It provides an expansion
\begin{equation}
\hat{U}^{I}\left(t\right)=e^{-i\sum_{n=0}^{\infty}\bar{H}_{n}(t)}\label{eq:Magnus_Expansion}
\end{equation}
thus truncating this series happens in the exponent and maintains
unitarity and is compatible with going to long times. Its lowest orders
can be understood as follows: We start with the average Hamiltonian
\[
\bar{H}_{0}(t)=\int_{0}^{t}d\tau\ \hat{V}^{I}(\tau)
\]
 i.e. the expression that collects the classical part and ignores
all commutators. The next order contains one commutator
\[
\bar{H}_{1}(t)=-\frac{i}{2}\int_{0}^{t}d\tau_{1}\,d\tau_{2}\ \left[\hat{V}^{I}(\tau_{2}),\hat{V}^{I}\left(\tau_{1}\right)\right]
\]
 but as it is in the exponent, it collects terms from all orders of
the Dyson series (you can convince yourself by expanding the exponential
in eq. (\ref{eq:Magnus_Expansion}). The next order of the expansion
is
\[
\bar{H}_{2}\left(t\right)=-\frac{1}{6}\int_{0}^{t}d\tau_{1}\,d\tau_{2}\,d\tau_{3}\,\left\{ \left[\hat{V}^{I}\left(\tau_{3}\right),\left[\hat{V}^{I}\left(\tau_{2}\right),\hat{V}^{I}\left(\tau_{1}\right)\right]\right]+\left[\hat{V}^{I}\left(\tau_{1}\right),\left[\hat{V}^{I}\left(\tau_{2}\right),\hat{V}^{I}\left(\tau_{3}\right)\right]\right]\right\}
\]
i.e. it contains two nested commutators . We will only be able to
appreciate this expansion when we apply it, but we can already see
that the different orders will inherit different operator structures
from the different commutators and that stacking on more integrals
will create ever more demanding resonance conditions, so higher orders
likely oscillate out. That notwithstanding, the Magnus expansion is
asymptotic in nature: Its formal radius of convergence is zero hence
adding higher orders does not always improve the accuracy.

\subsection{Real-world limitations}
When applying (quantum) optimal control to real-world systems, we have to contend with the fact that all parameters under our control have practical limitations: power, frequency, timing, etc. are all constrained by the capabilities of the equipment through which we apply said control. Moreover, any feedback scheme (such as Ad-HOC), must account for experimental noise, uncertainties in the experimental system (both gaps in system characterization, and "random walk"-like drifts of experimental parameters) and imperfections in both control and readout.

These issues above are complex and have to be dealt with simultaneously in real-word scenarios. There is no known textbook solution to these problems, and they are subject to ongoing research. We shall therefore limit ourselves to a very brief review of some of the approaches currently available:

\textbf{Constraints on applicable controls:} Two approaches can be taken: Either the space of possible controls can be defined such all points in the search space are valid, appplicable, controls, or the optimization space is defined more liberally, and we penalize controls which fail to conform.

For the first approach, limiting the control subspace, a partial solution is to choose and fix some parameters, such as control field frequency, ahead of time. This is the solution suggested by the CRAB optimal control algorithm \cite{CRAB}. A more general approach is to use bounded functions, such as cosine or inverse tangent, to transform an unconstrained physical parameter to a constrained one. For example, the search parameter $\alpha$ may be unconstrained and $\mathcal{O}\left(1\right)$, and we transform it to a constrained field amplitude via $A:=500\textrm{MHz}\times\cos\left(\alpha\right)$, which is subsequently used in the system Hamiltonian. 

Sometimes, the approaches above are insufficient as constraints are complex and include multiple parameters; or perhaps such substitutions are not a good fit to the optimal control problem. In such cases, we can impose a penalty term which will modify the functional for which we seek a minimum. For example, if we wish to impose a low-bandwidth solution on the control field $c\left(t\right)$, we may add a penalty term proportional to $\int_0^T {\left\vert \partial_t c\left(t\right)\right\vert}^2 dt$, which will be significant for highly oscillatory functions and zero for the DC component.

\textbf{Robust controls:} Experiments are often noisy environments, which noise appearing both on control fields and on the underlying system Hamiltonians. 
To provide a control scheme which provides consistently good performance, once must add the robustness requirement of the optimization requirements. 
This can be done using "ensemble optimization", where each optimization step averages over multiple manifestations of the dynamics, each with a different noise realization. 
The specific noise manifestations can be either fixed for the duration of the optimization of varied with each iteration step.
 The former approach is simpler to implement, but runs the risk of the optimization solving the problem only to the small subset of noises it encountered. 
 The latter approach tends to result in more robust controls, but introduces a noisy goal function, which is harder to optimize reliably. In either case, ensemble optimization tends to be expensive in terms of computational resources. 
 In some cases, it is possible to replace it with the a penalty term which is proportional to the absolute value of the gradient of the standard optimization goal with respect to the noisy variable (i.e. require that the control's performance will be weakly dependent on the noisy parameter). 
 In all cases, robust controls often exhibit the "no free lunch" rule of control theory -- robust controls often require more time, more bandwidth, or provide a worst average-case performance than their non-robust counterparts \cite{Khani12}.

\section{Examples}

\subsection{Optimal control of a qubit}

Let's start with a really elementary analytical example: A single
qubit with Hamiltonian $\hat{H}(t)=u(t)\hat{\sigma}_{x}$ looking
at the fastest state transfer possible from $|0\rangle$ to $e^{i\phi}|1\rangle$
. We can parameterize the state as $|\psi\rangle(t)=\left(x_{0}+iy_{0}\right)|0\rangle+\left(x_{1}+iy_{1}\right)|1\rangle.$
The Schrödinger equation can be expressed in these real parameters
as
\[
\dot{x}_{0}=uy_{1}\quad\dot{y}_{0}=-ux_{1}\quad\dot{x}_{1}=uy_{0}\quad\dot{y}_{1}=-ux_{0}
\]
 which are coupled in two sets of two that do not talk to the other
components, already telling us that $\phi=\pm\pi/2$. Keep in mind,
however, that $u$ can be time-dependent. Now we clearly see that
the speed of evolutions scales with the control amplitude $u$ so
our initial question was not even well-posed. We need to at least
limit the amplitude of the control field. We make this dimensionless
$\left|u\right|\le u_{{\rm max}}$. The optimal solution exhausts
that amplitude and, indeed, plugging in $u=u_{{\rm max}}$ we find
\[
\ddot{x}_{0}+u_{{\rm max}}^{2}x_{0}=0
\]
 the harmonic oscillator equation of motion which leads to the desired
solution $x_{0}=0$ after time $t_{{\rm min}}=\pi/2u_{max}$ . Solutions
of this kind are called ``bang'' solutions. More generally, in strictly
bilinear control problems like this one, the optimal solution jumps
between its boundaries (which in the case of multiple controls can
be quite intricate), then called ``bang-bang''-control.

It is interesting to study the physical significance of this result.
A real system in its laboratory frame always has an attached drift
\[
\hat{H}_{1}(t)=\frac{E}{2}\hat{\sigma}_{z}+u(t)\hat{\sigma}_{x}
\]
 Now if $u_{{\rm max}}\gg|E|$ we can expect the previous solution
to still hold approximately. If this condition is violated, the situation
is different: The vectors $\left(\pm u_{{\rm max}},y,z\right)^{T}$
define two non-collinear axes on the Bloch sphere and a given initial
state can reach all final states that are on the circle around that
axis including that state. In general, we will need up to three ``bangs''
to reach out goal. The limitation of $u_{{\rm max}}$ may
% A complete solution is described in the following article by U. Boscain:
% U. Boscain, P. Mason, ``Time Minimal Trajectories for a Spin 1/2 Particle in a Magnetic field '', J. Math. Phys. 47, 062101 (2006)
% The number of bang pulses can be computed as a function of u_{max}

\subsection{Exploring the speed limit with high parameter counts}

The \myiffindex{quantum speed limit} (QSL) is defined as the minimal time that
is needed to evolve a system from a given state $\rho_{0}$ to another
state $\rho(t)$ with a specific fidelity $\Phi(\rho_{0,}\rho(t))$
\cite{QSL0}. This is relevant e.g. for qubit gate implementations,
because it limits the minimal gate time (for unrestricted controls).
When the control bandwidth is restricted, then the dimension of the
set of reachable states $D_{\mathcal{{W}}}$ and the available bandwidth
$\Delta\Omega$ give a lower bound for the evolution time \cite{LLoyd14}:

\[
T\geq\frac{D_{\mathcal{{W}}}}{\Delta\Omega}
\]

This is a continuous version of the Solovay-Kitaev theorem.

The set of reachable states consists of all states that can be written
as

\begin{equation}
\ket{\psi(t)}=U(t_{0},t)\ket{\psi_{0}}
\end{equation}

where $U(t_{0},t)$ is the propagation operator of the system. A system
is called completely controllable if one can choose the control parameters
in such a way that the propagation operator is equal to any specific
operator \cite{Controllability}.

A method to explore the QSL for a gate is the following \cite{QSL2}:
For different given gate times one optimizes the gate and plots the
fidelity $\Phi_{\text{{goal}}}$ or the error $g(T)=1-\Phi_{\text{{goal}}}$
(see \eqref{goal}) of the optimized gates against the gate times.
If a QSL exists, there will be minimal time for which the error is
small. For shorter gate times the error is significantly larger. This
time is the QSL.

\begin{figure}[ht]\centering
	\includegraphics{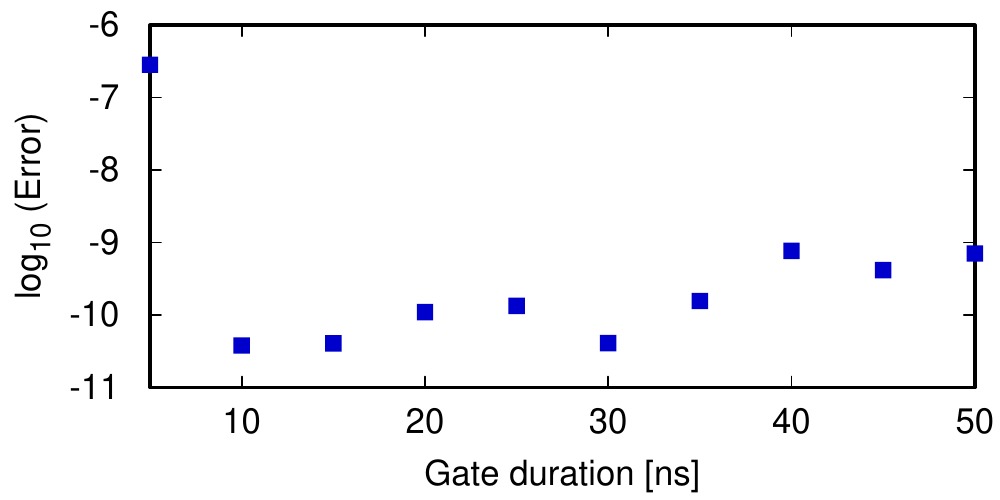}\caption{Gate error as a function of gate time. The optimization was done using
		GRAPE with a PWC parameterization with 500 pieces. The QSL is around
		$10$ns.}\label{fig:QSL_PWC}
\end{figure}

\begin{figure}[ht]\centering
	\includegraphics{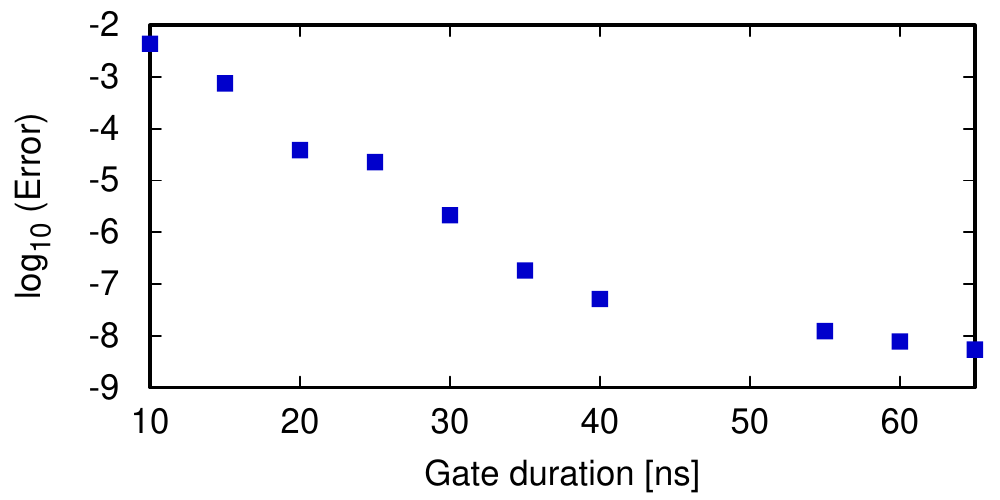}\caption{Gate error as a function of gate time. The optimization was done using
		GOAT with a Fourier decomposition into 167 pieces. The QSL is around
		$40$ns.}\label{fig:QSLFourier}
\end{figure}

The result depends on the chosen optimization method, concretely we show an example:

In fig. \ref{fig:QSL_PWC} and fig. \ref{fig:QSLFourier} the error $g$
is plotted against gate duration for two different parameterizations.
The system is a CR-gate implementation of a CNOT gate \cite{QSL3}.

In fig. \ref{fig:QSL_PWC} the QSL is shown for a piecewise constant
(PWC) parameterization with 500 pieces and unconstrained controls.
One can see that there is a jump around 10ns which indicates that
this is the QSL in this case. Fig. \ref{fig:QSLFourier} shows the
same, but with a Fourier decomposition into 167 components. The QSL
is here around 40ns and is reached more slowly.

The difference is related to optimally have the controls interact with redirecting the drift. A key step to a theory of this phenomenon has been undertaken in \cite{LLoyd14}.

\subsection{Open systems}

In these notes we have mostly concentrated on the optimal control
for closed quantum systems. One can ask related questions for open
quantum systems as well. A treatment of this situation would go way
beyond the scope of these lecture notes. Here, the space of potentially
reachable states / of reachable time evolutions is much larger than
in the unitary case. The theory of controllability and reachability is
thus more involves, it is for example not at all clear, if the impact
of decoherence can be reduced to zero, i.e., if the subset of unitary
time evolutions is reachable. We would thus like to describe a pragmatic
approach and refer the reader to the literature.
%DS
For a Lindblad equation, it can be shown that the control fields cannot cancel dissipation effect and the system is not completely controllable.
This is still an open question in the non-Markovian regime.

As a first rule of thumb, there are situations when the decoherence
experienced by the quantum subsystem has no or very little structure
-- e.g. in the case of uniform decoherence leading to a fully depolarizing
channel and, at least for the synthesis of gates, for most Markovian
decoherence models. These do not give an open system optimal control
algorithm any space to actually exploit the structure of the decoherence
to perform an optimization, rather, we can expect that the fastest
solution of the closed system also is close to an optimal solution
for the open system. Thus, running a closed-system version of optimal
control and benchmarking it in a realistic open system is a good initial
approach.

If one suspects that the decoherence mechanism contains exploitable
structure, or if one tries to accomplish a task that actively uses
decoherence -- such as tasks changing the entropy of the state, e.g.,
cooling, it is possible to generalize the aforementioned methods of
optimal control. More specifically, e.g., in \myiffindex{OpenGRAPE}, one simply
replaces the Schrödinger equation as the dynamical constraint by a
suitable description of open systems dynamics, such as a master equation.
One caveat lies in the need for backwards-in-time propagation: Open
system dynamics is asymptotically irreversible, which can make back-propagation
unstable. Practically, this can be handled by either focusing on decoherence
rates that are not too large or by suitable initial guesses.

As a well-defined example, let us consider a single qubits perturbed
by a two-level fluctuator, i.e., a second two-state system that is
coupled to a heat bath. This is a common situation in superconducting
qubits \cite{Rebentrost07}.

We specifically model a qubit coupled to a single TLF by ${H}={H}_{S}+{H}_{I}+{H}_{B}$. ${H}_{S}$
consists of the qubit and the coupled two-state system, i.e.
\[
{H}_{S}=E_{1}(t){\sigma}_{z}+\Delta{\sigma}_{x}+E_{2}{\tau}_{z}+\Lambda{\sigma}_{z}{\tau}_{z}
\]
 where ${\sigma}_{i}$ and ${\tau}_{i}$ are the usual Pauli matrices
operating in qubit and fluctuator Hilbert space respectively. $E_{1}(t)$
is time-dependent and serves as an external control. The source of
decoherence is the coupling of the fluctuator to the heat bath, which
leads to incoherent transitions between the fluctuator eigenstates,
${H}_{I}=\sum_{i}\lambda_{i}({\tau}^{+}{b}_{i}+{\tau}^{-}{b}_{i}^{\dagger}),\quad{H}_{B}=\sum_{i}\hbar\omega_{i}b_{i}^{\dagger}b_{i}.$
We introduce an Ohmic bath spectrum $J(\omega)=\sum_{i}\lambda_{i}^{2}\delta(\omega-\omega_{i})=\kappa\omega\Theta(\omega-\omega_{c})$
containing the couplings $\lambda_{i}$, the dimensionless damping
$\kappa$, and a high-frequency cutoff $\omega_{c}$ (which we assume
to be the largest frequency in the system). Now depending on the bath
damping constant $\kappa$the fluctuator can flip fast or slow -- and
in the limit of slow flipping, the qubit sees noise with strong temporal
correlation leading to highly non-Markovian qubit dynamics.

To formally treat this system, we can on the other hand still set
up a Markovian master equation for the augmented system of qubit \emph{and
} fluctuator and only after its solution trace over the fluctuator
to get the effective density matrix of the qubit alone. We formulate
the control approach by rewriting the master equation as $\dot{\rho}(t)=-\big(i\mathcal{H}(E_{1}(t))+\Gamma(E_{1}(t))\big)\rho(t)$
with the Hamiltonian commutator superoperator $\mathcal{H}(E_{1}(t))(\cdot)=[H(E_{1}(t)),\cdot]$
and the relaxation superoperator $\Gamma$, both time-dependent via
the control $E_{1}(t)$. The formal solution to the master equation
is a linear quantum map operating on a physical initial state according
to $\rho(t)=F(t)\rho(0)$. Thus $F$ itself follows the operator equation
of motion
\begin{equation}
\dot{F}=-\left(i\mathcal{H}+\Gamma\right)F\label{eq:FTimeEvolution}
\end{equation}
with initial condition $F(0)=\mathbb{I}$, as in ref. \cite{Tosh06}.
%DS:
%We have recently studied this problem with Christiane from a numerical and analytical points of view. We explore the time minimum way to purify the qubit. Maybe we could cite the two references:
%Beating the limits with initial correlations
%D. Basilewitsch, R. Schmidt, D. Sugny, S. Maniscalco and C. P. Koch
%New J. Phys. 19, 113042 (2017)
%Time-optimal control of qubit purification in contact with a structured environment
%J. Fisher, D. Basilewitsch, C. P. Koch and and D. Sugny
%Phys. Rev. A 99, 033410 (2019)

Here, multiplication of quantum maps denotes their concatenation.
The task is to find control amplitudes $E_{1}(t)$ with \$$t\in[0,t_{g}]$\$,
\$$t_{g}$\$ being a fixed final time, such that the difference \$$\delta F=F_{U}-F(t_{g})$\$
between dissipative time evolution \$$F(t_{g})$\$ obeying eqn. (\ref{eq:FTimeEvolution})
and a target unitary map \$$F_{U}$\$ is minimized with respect to
the Euclidean distance $||\delta F||_{2}^{2}\equiv{\rm tr}\left\{ \delta F^{\dagger}\delta F\right\} $.
Clearly, this is the case, when the trace fidelity
\begin{equation}
\phi={\rm {Re\,tr}\left\{ F_{U}^{\dagger}\;F(t_{g})\right\} }\label{eq:FidelityOpenSystem}
\end{equation}
is maximal. Note, that in an open system, one cannot expect to achieve
zero distance to a unitary evolution $F_{U}$\cite{Tosh06}.
The goal is to come as close as possible. On this setting, we find
optimal pulses by gradient search.

It is interesting to investigate the resulting pulses and performance
limits. We see in figure ... that optimal control pulses allow to
reach great gate performance after overcoming a quantum speed limit.
Remarkably, the dependence on gate duration is non-monotonic at least
in the regime of low $\kappa$ when the two settings of the TLS can
be resolved. At some magic times, the frequency split from the TLS
naturally refocuses, constraining the optimization much less than
at other times.

\begin{figure}\centering
\includegraphics[width=0.9\linewidth]{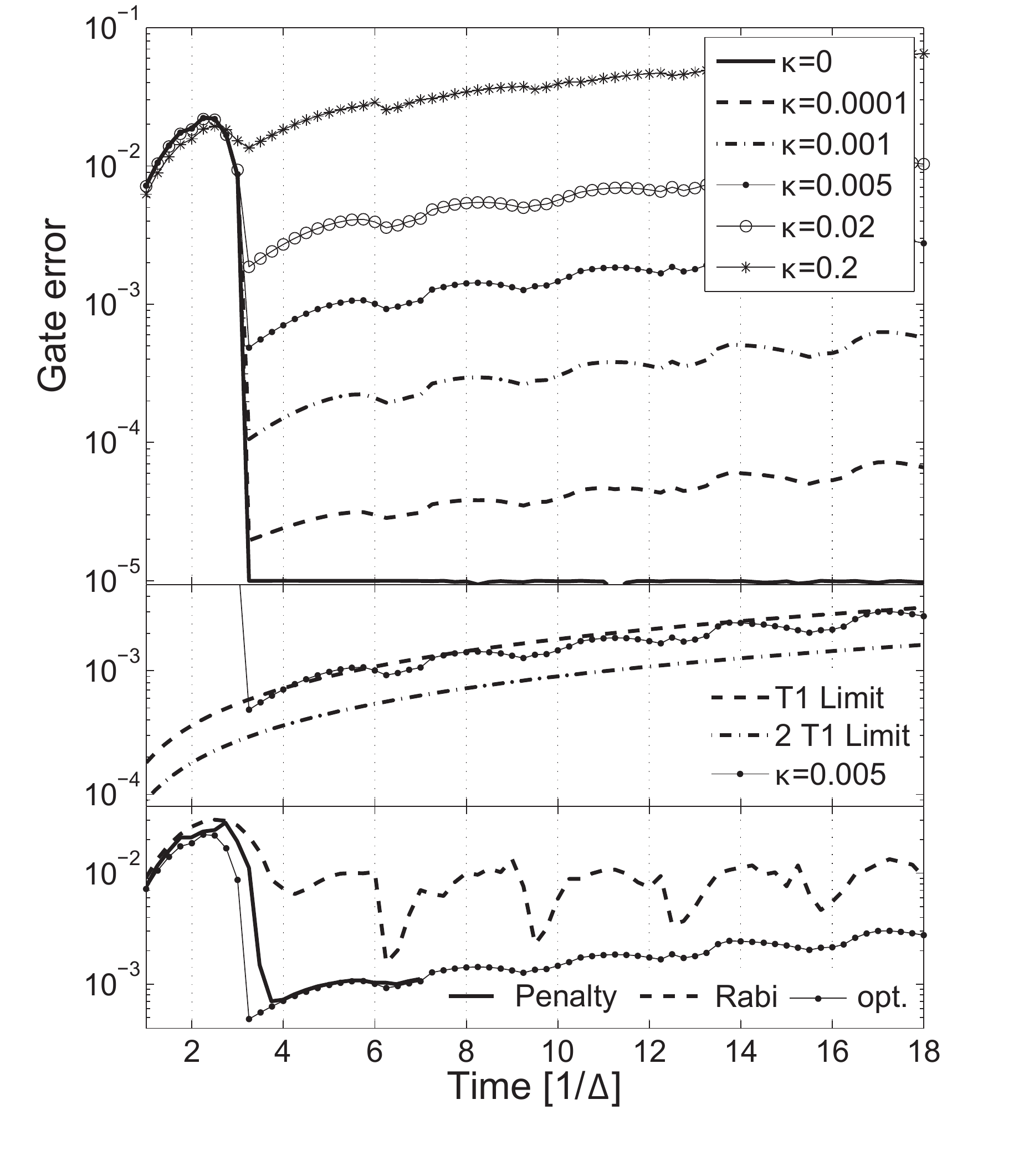}

\caption{Top: Gate error versus pulse time $t_{g}$ for optimal Z-gate pulses
in the presence of a non-Markovian environment with dissipation strength
$\kappa$. A periodic sequence of minima at around $t_{n}=n\pi/\Delta$,
where $n\ge1$, is obtained. Middle: The gate error of optimized pulses
approaches a limit set by $T_{1}$ and $2T_{1}$, as shown with $\kappa=0.005$.
Bottom: Optimized pulses reduce the error rate by approximately one
order of magnitude compared to Rabi pulses for $\kappa=0.005$. Pulses
starting from zero bias and with realistic rise times (penalty) require
only a small additional gate time. In all figures the system parameters
are $E_{2}=0.1\Delta$, $\Lambda=0.1\Delta$ and $T=0.2\Delta$.}
\end{figure}

More remarkable, the maximally attainable fidelity also has a non-monotonic
dependence on $\kappa$. At hindsight, this can be understood as follows:
At low $\kappa$ there is no randomness of the system, it is fully
reversible. The optimal control algorithm just has to deal with the
fact that the setting of the TLS is unknown, which it perfectly accomplishes.
On the other hand, at high $\kappa$, the phenomenon of \emph{motional
narrowing }occurs: Fast motion of the impurity broadens its spectrum
thus reducing its spectral weight at low frequencies.

\begin{figure}\centering
\includegraphics[width=0.9\columnwidth]{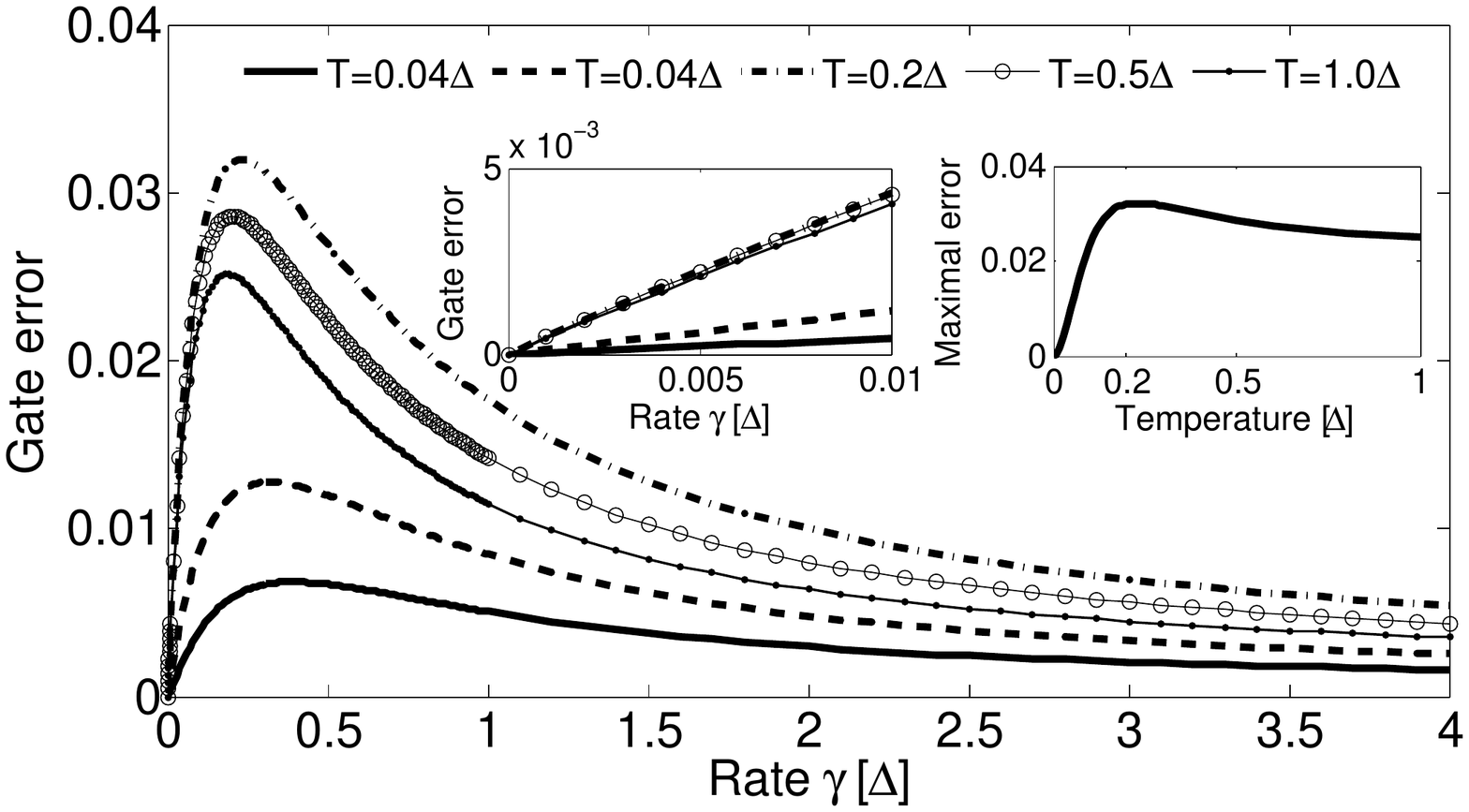}

\caption{Gate error versus TLF rate $\gamma$ for various temperatures for
an optimized pulse with $t_{g}=5.0/\Delta$. The left inset is a magnification
of the low-$\gamma$ part of the main plot and reveals the linear
behaviour. The right inset shows the maximum of the curves of the
main plot versus temperature. ($E_{2}=0.1\Delta$ and $\Lambda=0.1\Delta$}
\end{figure}

\subsection{DRAG and its derivatives}

In general a quantum system will contain additional states outside
of a specific subspace we want to operate in. If our control couples also
to transitions out of the subspace we will leak population and degrade
the performance of our operation. The \myiffindex{Derivative Removal with Adiabatic Gate (DRAG) }method provides a framework
to identify these leakages and to modify the control signals to counteract
them.

We will review the basic idea along the procedure shown in \cite{Motzoi2009}.
Consider a 3-level-system that is controlled by a signal $u(t)=u_{x}(t)\cos(\omega_{d}t)+u_{y}(t)\sin(\omega_{d}t)$.
The first two levels make up the computational subspace $\ket 0,\ket 1$
with transition frequency $\omega_{1}$ that we want to operate in
and $\ket 2$ accounts for the leakage. It is modeled by the Hamiltonian

\begin{equation}
H/\hbar=\omega_{1}\ketbra 11+(2\omega_{1}+\Delta)\ketbra 22+u(t)\hat{\sigma}_{0,1}^{x}+\lambda u(t)\hat{\sigma}_{1,2}^{x}\label{eq:ham_qutrits}
\end{equation}

where the Pauli operators are $\hat{\sigma}_{j,k}^{x}=\ketbra jk+\ketbra kj$ and
$\lambda$ describes the coupling of the drive to the 1-2 transition.
We expressed the second transition frequency by the anharmonicity
$\Delta=\omega_{2}-2\omega_{1}$.

Let's say we want to implement a simple Rabi pulse by choosing $u_{x}(t)=\Omega(t)$
and $u_{y}(t)=0$. This gives rise to unwanted leakage out of the
computational subspace with the term $\lambda\Omega(t)\hat{\sigma}_{1,2}^{x}$.
The DRAG idea shows how we can counteract this leakage by choosing
$u_{y}(t)$ appropriately.

We first express the Hamiltonian in the rotating frame with $R=\exp(i\omega_{d}\ketbra 11+2i\omega_{d}\ketbra 11)$
following the rule $H^{R}=RHR^{\dagger}+i\hbar\dot{R}R^{\dagger}$
which gives

\[
H^{R}/\hbar=\delta_{1}\ketbra 11+\delta_{2}\ketbra 22+\sum_{\alpha=x,y}\frac{u_{\alpha}}{2}(t)\hat{\sigma}_{0,1}^{\alpha}+\lambda\frac{u_{\alpha}}{2}(t)\hat{\sigma}_{1,2}^{\alpha}\:,
\]
using the detunings $\delta_{1}=\omega_{1}-\omega_{d}$ and $\delta_{2}=\Delta+2\delta_{1}$
between the drive and transition frequencies.

Applying an adiabatic transformation $V(t)$ by calculating $H^{V}=VHV^{\dagger}+i\hbar\dot{V}V^{\dagger}$
allows us to look at the system in a frame where the leakage and the
$y$-component necessary to counteract it are visible. We take
\[
V(t)=\exp\left[-i\frac{u_{x}(t)}{2\Delta}(\hat{\sigma}_{0,1}^{y}+\lambda\hat{\sigma}_{1,2}^{y})\right],
\]

a transformation that depends on our intended signal $u_{x}$, and
apply it to first order in $u_{x}/\Delta$ to find
\[
\begin{aligned}H^{V}/\hbar & =\left(\delta_{1}-\frac{(\lambda^{2}-4)u_{x}^{2}}{4\Delta}\right)\ketbra 11+\left(\delta_{2}+\frac{(\lambda^{2}+2)u_{x}^{2}}{4\Delta}\right)\ketbra 22\\
 & +\frac{u_{x}}{2}\hat{\sigma}_{0,1}^{x}+\lambda\frac{u_{x}^{2}}{8\Delta}\hat{\sigma}_{0,2}^{x}+\left[\frac{u_{y}}{2}+\frac{\dot{u}_{x}}{2\Delta}\right](\hat{\sigma}_{0,1}^{y}+\lambda\hat{\sigma}_{1,2}^{y})
\end{aligned}
\]

From this expression we can see that our intended drive is unchanged
$u_{x}/2\hat{\sigma}_{0,1}^{x}$ but if we also choose $u_{y}=-\dot{u}_{x}/\Delta$
we cancel the last term that is responsible for driving out of the
computational subspace $\propto\lambda\hat{\sigma}_{1,2}^{y}$. The
transformation also suggest detuning the drive by $\delta_{1}=(\lambda^{2}-4)u_{x}^{2}/4\Delta$
to avoid stark shifting of the 0-1 transition. This example illustrates
the main working principle of DRAG which can be generalized to account
for more than just leakage to a third level. By modifying $V(t)$,
for example adding terms $\propto\hat{\sigma}_{0,2}^{y}$, or iteratively
performing transformations $V_{j}(t)$ the intertial terms, the inertial
terms $i\hbar\dot{V}_{j}V_{j}^{\dagger}$ generate more conditions
on the control signals and its derivatives.

\begin{figure}\centering
\includegraphics[width=0.8\textwidth]{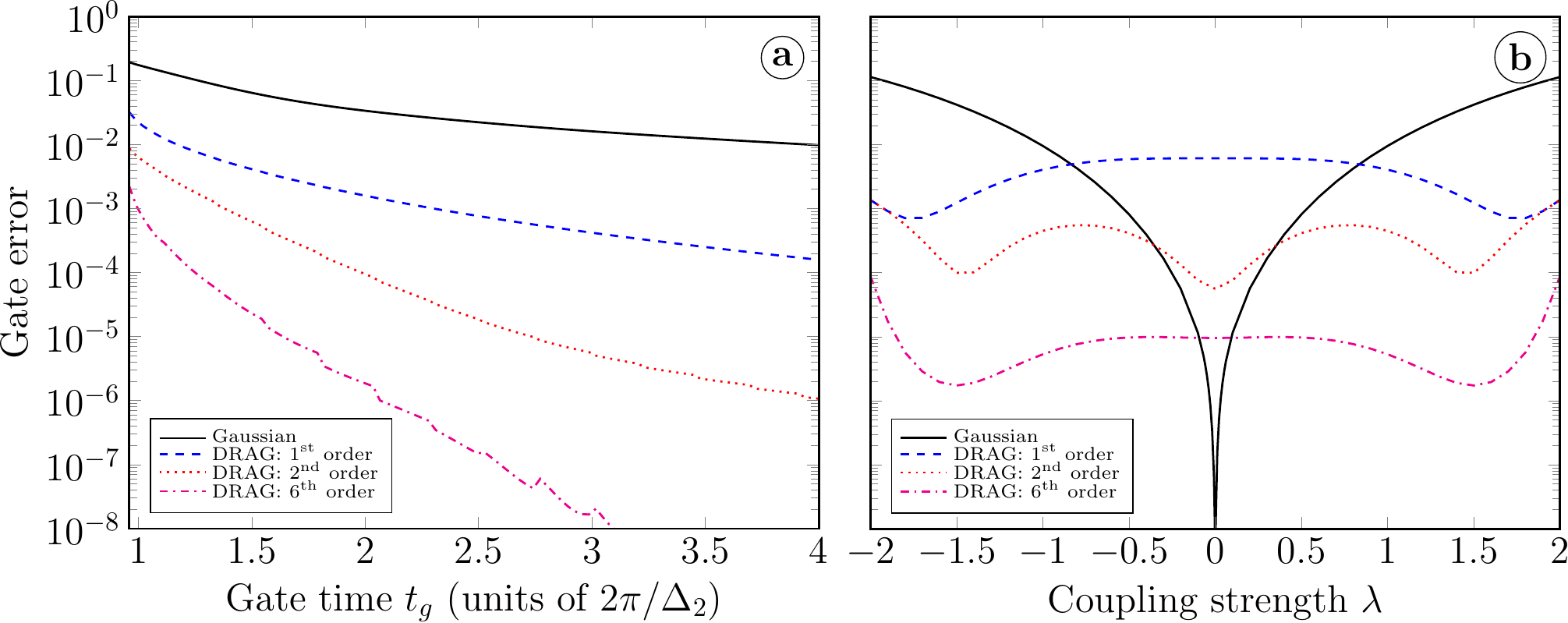}
\caption[Gate error for analytic DRAG control to different orders.]{\label{fig:DRAG_perform}(a): Performance of non-optimized DRAG variants
as a function of gate time, derived from an iterative Schrieffer-Wolff
expansion to higher orders. Target : $\hat\sigma_{x}$ rotation of
a single qubit described by the lowest three levels of Hamiltonian
(\ref{eq:ham_qutrits}). -- (b): Performance of the DRAG pulses used in (a) for a fixed gate time $t_{g}=4\pi/\Delta_{2}$ as a function of coupling strength $\lambda$ to the leakage level.}
\end{figure}

The performance of solutions to different orders, obtained via iterative
transformations, is depicted in Fig.\ref{fig:DRAG_perform}a as a
function of pulse length, and in Fig.\ref{fig:DRAG_perform}b as a
function of coupling strength $\lambda$ for a fixed gate time $t_{g}=4\pi/\Delta_{2}$.
Higher order solutions are taken from \cite{Motzoi2013}. Note also
that when the $\ket 0\leftrightarrow\ket 2$ transition is controlled
via an additional corresponding frequency component, exact solutions
to the three-level system exist (cf.~chapter 8 in \cite{Motzoi2012}).

Turning to the experimental implementation \cite{Lucero2010,Chow2010}
of DRAG pulses: In practice, actual system parameters differ somewhat
from those assumed in theory due to characterization gaps, system
drift, or unknown transfer functions affecting the input field shapes
\cite{Motzoi2011}. As a simplification, we assume the low order terms
in DRAG are easier to implement as their shape will be mostly maintained
on entry into the dilution fridge. Even so, many different low-order
variants of DRAG have been found in the literature for third-level
leakage \cite{Motzoi2009,Gambetta2011a,Motzoi2013,Lucero2010}. This
reduced functional form can further be optimized theoretically \cite{Theis2016}
and/or through a closed-loop process experimentally \cite{AdHOC,ORBIT}
to account for the effect of higher order terms and experimental uncertainties
(preferably using more advanced gradient-free algorithms such as CMA-ES
\cite{Hansen2003}). A systematic experimental study of the tune-up
of the prefactors in front of the functional forms for the control
operators was performed in \cite{Chen2016}. In writing up these optimizations and adapting them, the Magnus expansion, see chapter \ref{ch:Magnus} is typically used.

\begin{figure}[t]
\centering \includegraphics[width=0.8\linewidth]{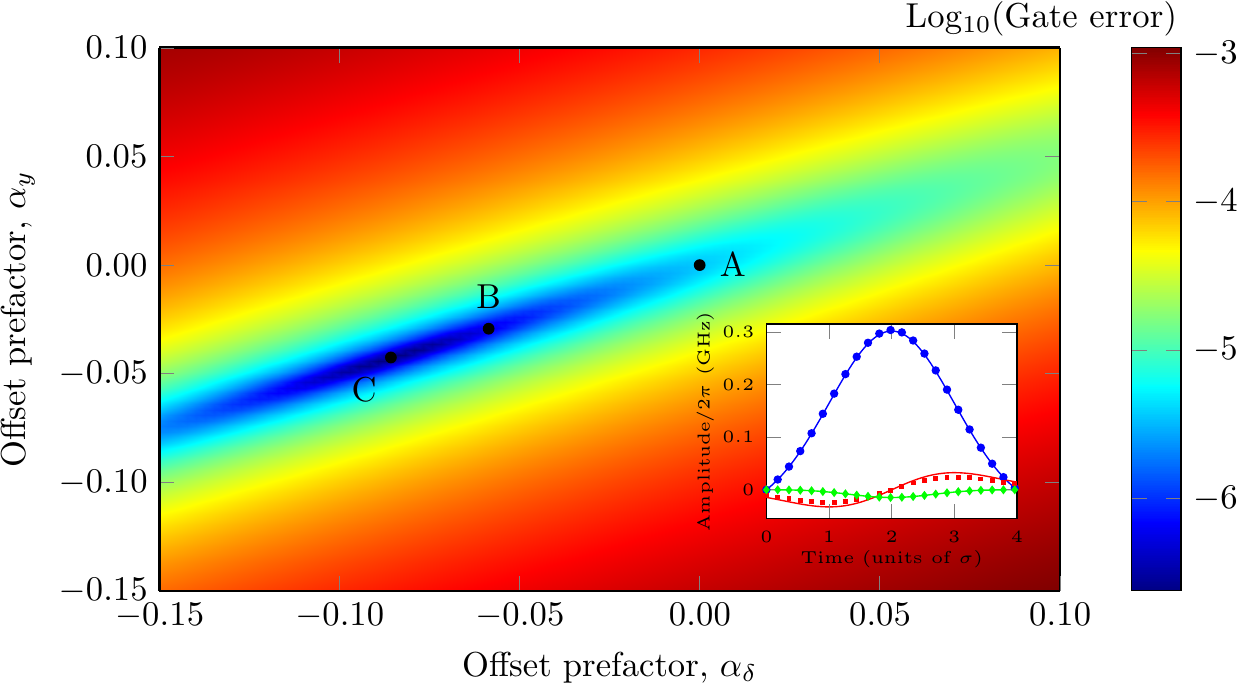}
\caption[Calibration landscape for first order DRAG solutions to the single-qubit
leakage problem.]{\label{fig:calibration-landscape}A slice of the 3D calibration landscape
for DRAG solution up to the first order in the small parameter to
the qubit $\sigma_{x}$-gate leakage problem. Point A and B denote
\cite{Motzoi2009}'s and \cite{Gambetta2011a}'s first-order solutions,
respectively. Point C is the optimum for this control function subspace
(here $\alpha_{x}=-0.0069$), with infidelity of $10^{-6.63}$. A
successful calibration process will typically start at a known DRAG
solution, i.e. points A or B, and conclude in point C. The inset illustrates
the associated pulse shapes: markers represent the unoptimized shapes
($u_{x}$: ${\color{blue}\bullet}$, $u_{y}$: ${\color{red}\blacksquare}$,
$\delta$: ${\color{green}{\color{lime}{\color{green}\blacklozenge}}}$)
whereas solid lines depict the corresponding optimal solution (C).}
\end{figure}

For instance, let us denote the Gaussian pulse implementing a $\hat{\sigma}_{x}$
gate for the qubit by $G(t)$. Then the first order solutions described
in \cite{Motzoi2009,Gambetta2011a,Motzoi2013} are parameterized by
the limited functional basis $u_{x}\propto G$, $u_{y}\propto\partial_{t}G$
and $\delta\propto G^{2}$, which mimics the limited shaping control
that can exist in experiment. None of the reported solutions are optimal
within this functional basis: For typical example parameters, infidelities
may be further reduced from $10^{-5.28}$ to $10^{-6.63}$ by slightly
adjusting the prefactors of the control fields. For example, \cite{Motzoi2009}'s
first order DRAG solution may be transformed according to $u_{x}\to(1+\alpha_{x})u_{x}$
and similarly for $u_{y}$ and $\delta$, and then the constants $\alpha_{x}$,
$\alpha_{y}$ and $\alpha_{\delta}$ are optimized. A discussion for
why optimization within a severely restricted functional subspace
may often be sufficient is given in \cite{Caneva2011} and follow-up
publications. A schematic of the optimization task involved in
the calibration, as well as the shape of the associated controls,
is shown in Fig.\ref{fig:calibration-landscape}.

\section{Summary and outlook}

Optimal control is a mature discipline of theoretical physics and related fields. In experimentation, it has remarkable success in situations in which physical systems are well characterized. Reaching out to engineered systems requires a close integration with characterization and benchmarking.

Experimentalists and users of quantum control should have taken home an introduction of concepts, jargon, and results of the field. Theorists should feel motivated to embrace these challenges and to fashion their results into tools that can be used efficiently and scalably so quantum control and quantum technology applications can mutually benefit from their potential.

\section*{Acknowledgements}

We acknowledge collaboration with the optimal control group at Saarland
University (and its previous locations), including Daniel Egger, Likun Hu, Kevin Pack, Federico Roy, Ioana Serban, and  Lukas
Theis  as well as continuous collaboration with Tommaso Calarco,
Steffen Glaser, Christiane Koch, Simone Montangero, and Thomas Schulte-Herbrüggen. Some
of this work is sponsored by the Intelligence Advanced Research Projects
Activity (IARPA) through the LogiQ Grant No. W911NF-16-1-0114, by
the European Union under OpenSuperQ and the ITN Qusco.

%%%%%%%%%%%%%%% References %%%%%%%%%%%%%%%%%
\addcontentsline{toc}{section}{References}  % puts references starting page into contents as in MS Word template

\bibliography{bibliography, Lukas_Bibliography}

% For manual generation of references with correct format uncomment this:
%\begin{thebibliography}{99}
%\frenchspacing
%\bookref{Sq78}{G. L. Squires}{Introduction to the theory of thermal neutron scattering}{Cambridge University Press}{Cambridge}{1978}
%\bibitem{Br08} T. Br\"uckel, G. Heger, D. Richter, R. Zorn (eds.), {\it Neutron Scattering} (Forschungszentrum J\"ulich, 2008), \url{http://hdl.handle.net/2128/3718}
%\bookref{Lo84}{S. W. Lovesey}{Theory of Neutron Scattering from Condensed Matter}{Clarendon Press}{Oxford}{1984}
%\jref{vM98}{W. van Megen, T. C. Mortensen, S. R. Williams, J. M\"uller}{Phys. Rev. E}{58}{6073}{1998}
%\end{thebibliography}

\end{document}